\documentclass[graphics, twocolumn, usenatbib]{mn2e}
\usepackage{times}
\usepackage{mathptmx}
\usepackage{epsfig} 
\usepackage{graphicx} 
\usepackage{color}
\usepackage{deluxetable} 
\usepackage{aas_macros} 
\usepackage{amssymb}
\usepackage{amsmath}
\usepackage{yhmath}
\usepackage[title]{appendix}

\newcommand{\enzo}{\textsc{Enzo~}}
\newcommand{\enzolat}{\textsc{Enzo-2.3}}
\newcommand{\cosmos}{\textsc{Cosmos~}}

\newcommand{\kms} {km $\rm{s^{-1}}$}
 
\newcommand{\msolar} {$M_{\odot}~$}
\newcommand{\msolarc} {$M_{\odot}$}
\newcommand{\molH} {$\rm{H_2}$~}

\def\etal{{\it et al.}~}

\title[]{Numerical resolution effects on simulations of massive black hole seeds}

\author[J.A. Regan \etal] 
{John A. Regan$^{1,2}$\thanks{E-mail:john.regan@helsinki.fi}, Peter H. Johansson$^{1}$ 
\& Martin G. Haehnelt$^{2}$ \\ \\
$^1$ Department of Physics, University of Helsinki, Gustaf H\"allstr\"omin katu 2a,
FI-00014 Helsinki, Finland \\
$^2$ Kavli Institute for Cosmology and Institute of Astronomy, Madingley Road, 
Cambridge CB3 0HA, United Kingdom
\\}

\begin{document}

\maketitle

\begin{abstract} 
We have performed high-resolution numerical simulations with the hydrodynamical AMR code 
\enzo to investigate the formation of massive seed black holes in a 
sample of six dark matter haloes above the atomic cooling threshold. The aim of this study is 
to illustrate the effects of varying the maximum refinement level on the final object formed. 
The virial temperatures of the simulated haloes range from $T \sim 10000\ \rm{K} - 16000$ K and 
they have virial masses in the range $M \sim 2 \times 10^7$\msolar \ to $M \sim 7 \times 
10^7$\msolar at $z \sim 15$.  The outcome of our six fiducial simulations is both generic and 
robust. A rotationally supported, marginally gravitationally stable, disk forms with an exponential 
profile. The mass and scale length of this disk depends strongly on the maximum refinement level 
used. Varying the maximum refinement level by factors between $ {1 / 64}\ \rm{to} \ {256}$ times the 
fiducial level illustrates the care that must be taken in interpreting the results. 
The lower resolution simulations show tentative evidence that the gas may become rotationally 
supported out to 20 pc while the highest resolution simulations show only weak evidence of 
rotational support due to the shorter dynamical times for which the simulation runs. The higher 
resolution simulations do, however, point to fragmentation at small scales of the order of 
$\sim 100 \ \rm{AU}$. In the highest resolution simulations a central object of a few times 
$10^2$\msolar forms with multiple strongly bound, Jeans unstable, clumps of $\approx 10$ \msolar and 
radii of 10 - 20 AU suggesting the formation of dense star clusters in these haloes.

\end{abstract}



\section{Introduction} 
It is now widely accepted that supermassive black holes (SMBHs)  populate the centres of 
most if not all galaxies.  SMBHs were invoked early on in the literature to explain
the powering of extremely luminous, extra-galactic sources \cite[]{Lynden-Bell_1969} initially 
dubbed quasi-stellar objects (QSOs) \cite[]{Zeldovich_1964, Salpeter_1964}. Since their 
discovery nearly five decades ago much theoretical and observational work has gone into 
explaining both their presence and their properties. The mass of central SMBHs appears
to correlate surprisingly strongly  with the luminosity and the stellar velocity 
dispersion of the galactic bulges hosting them \citep[see \citealt{Kormendy_2013} for a recent
review]{Magorrian_1998,Gebhardt_2000, Ferrarese_2000}. The physical origin of the claimed 
tight correlation of stellar velocity dispersion and black hole mass, the ``$M_{\rm BH}- \sigma$'' 
relation, is still controversial (see \citealt{Fabian_2012} for a recent review), but is generally 
attributed to the joint formation history of (proto-)galaxies and their central black hole 
\citep{Kauffmann_2000,Haehnelt_2000}. Numerical simulations that include  models for the feedback 
from the SMBHs on the surrounding gas support this picture (e.g. \citealt{DiMatteo_2005, 
Sijacki_2006, Sijacki_2007, Sijacki_2009,  Johansson_2009b, Johansson_2009a, Debuhr_2011, 
Choi_2013}). \\ \indent
Surprisingly, billion solar mass black holes  already exist  at $z \ga 6$ when the Universe was
less than a Gyr old \citep{Fan_2004, Fan_2006, Mortlock_2011,Venemans_2013}. This poses 
problems for models that assume Eddington limited growth of stellar mass  seed black holes 
in the limited time available (e.g. \citealt{Costa_2013} and see \citealt{Volonteri_2010a, 
Haiman_2012} for recent reviews). Stellar mass seed black holes form in shallow potential 
wells and their growth is hampered by the (negative) feedback due to photo-ionisation heating 
and the energy and momentum injection due to supernovae \citep{Haiman_2000, Haiman_2003, 
Johnson_2007, Greif_2007, Bromm_2009, Pawlik_2009}. The problems of models invoking Eddington 
limited growth have been further compounded by recent simulations reporting more efficient 
fragmentation than earlier work lowering the expected masses of population III (POP III) stars 
to $\approx 10 M_{\odot} - 100 M_{\odot}$  \cite[]{Turk_2009, Clark_2011, Stacy_2012, Greif_2012}.  
This increased fragmentation is due to a combination of turbulence,  H$_{2}$ collisional
dissociation cooling and collision-induced emission which could not be followed in earlier 
simulations due to resolution constraints. \\ \indent
Growth from massive seed black holes with masses above $10^4$ \msolar that form in a rapid 
phase where the Eddington limit is exceeded by a large factor is therefore often invoked as 
an alternative  \cite[]{Begelman_1978, Begelman_1984, Loeb_1994, Haiman_2001, Begelman_2001, 
Oh_2002, Bromm_2003}. The most promising route for the formation of such massive seed black
holes is probably ``direct collapse'' in dark matter haloes with virial 
temperatures above the atomic cooling threshold in which cooling  by molecular 
hydrogen and metals is not important and fragmentation is therefore 
suppressed \citep{Shlosman_1989, Loeb_1994, Begelman_2008, Begelman_2008b, Begelman_2006, 
Volonteri_2010}. A number of studies has shown that it may indeed be plausible that a 
sufficient number of such haloes has not been enriched by metals and are subject to 
sufficiently strong (local) UV radiation that molecular hydrogen is dissociated \citep{Bromm_2003, 
Haiman_2006, Haiman_2006b, Mesinger_2006,Dijkstra_2008, Cen_2008, Ahn_2009, Shang_2010, 
Wolcott-Green_2011, Tanaka_2013}. Note, however, recent work by \cite{Aykutalp_2013} on 
(negative) photo-ionisation feedback driven by X-rays has also shown that the composition of the 
gas surrounding the seed black hole may play a role in determining the final outcome.\\
\indent Simulations of the isothermal collapse of the gas in such haloes 
are numerically challenging, but significant progress has been made
mainly, but not exclusively, with grid based adaptive mesh refinement (AMR) codes 
(e.g. \citealp{Johnson_2007, Wise_2008, Regan_2009, Greif_2008, Johnson_2011, Latif_2012, 
Latif_2013a, Latif_2013b, Latif_2013c, Latif_2013d, Prieto_2013, Latif_2013e}). 
In the simulations the gas cools efficiently to temperatures of about 7000-8000K and becomes 
strongly turbulent. Turbulent angular momentum transport thereby leads to substantial and
sustained mass inflow and allows the gas to settle into an isothermal 
($\rho \propto r^{-2}$) density profile. A major difficulty with 
these simulations thereby stems from the fact that the dynamical timescales get shorter as the 
collapse progresses and the simulations are generally able to follow only an increasingly smaller 
fraction of the gas as it collapses to the highest densities. This makes it very difficult 
to follow gas that has settled into angular momentum support further out in the halo while the 
inner part continues to collapse. This has led to discussions as to the extent of angular momentum 
support and the subsequent fragmentation scale \citep{Wise_2008, Regan_2009, Latif_2013c}.  \\
Making progress with these questions is important for judging the
the importance of the ``direct collapse'' model as a route to the
formation of supermassive black holes. We therefore follow on here from our previous studies of 
this problem with a larger suite of simulations performed with increased resolution with a newer 
version of the \enzo code. In this way we are able to follow the collapse to higher density as well 
studying to an unprecedented extent the longer term evolution of (marginally)
angular momentum supported gas during the collapse. The larger sample of simulations also 
allowed us to look for systematic trends in the environmental properties of the haloes. 
The aim being to search for clues within the large scale environment which may favour the 
formation of massive seed black holes. 

The paper is structured as follows. In \S \ref{Sims} we describe the
details of the numerical  simulations. In \S \ref{results},
appendix \ref{appendixA} \& \S \ref{Scaling} we describe the results of
our numerical simulations and in \S \ref{conclusions} we deliver our conclusions.
Throughout this paper we  assume a standard $\Lambda$CDM cosmology with the following parameters 
\cite[based on the latest Planck data]{Planck_2013a}, $\Omega_{\Lambda,0}$  = 0.6817, 
$\Omega_{\rm m,0}$ = 0.3183, $\Omega_{\rm b}$ = 0.0463, $\sigma_8$ = 0.8347 and $h$ = 0.6704. 
We further assume a spectral index of primordial density fluctuations of $n=1$.

\begin{table*}
\centering
\begin{minipage}{160mm}
\begin{tabular}{ | l | l | l | l | l | l | l | l | l | l | l | l }
\hline \hline 
\em{Sim}$^a$
& \textbf{\em $z_{init}$$^{b}$} & \textbf{\em $Boxsize$$^{c}$} 
& \textbf{\em $z_{end}$$^{d}$} & \textbf{\em $M_{200}$$^{e}$}  & \textbf{\em $R_{200}$$^{f}$}
& \textbf{\em $V_{200}$$^g$} & \textbf{\em $T_{vir}$$^{h}$}  & \textbf{\em $M_{DM}$$^{i}$} 
& \textbf{\em $n_{max}$$^{j}$} & \textbf{\em $T_{core}$$^{k}$} & \em{$\Delta$ R$^{l}$} \\ 
\hline 
A & 100.0 & 2.0 & 21.8844 & 2.61 $\times 10^7$ & 0.402 & 16.65 &  $9979 $ & $2.11 \times 10^7$&  
$3.75 \times 10^{11}$ & 6842 & $3.88 \times 10^{-03}$  \\
B & 100.0 & 2.0 & 15.9407 & 3.57 $\times 10^7$ & 0.605 & 15.95 & $9157 $ & $2.95 \times 10^7$ &  
$2.77 \times 10^{11}$ & 7383 & $5.24 \times 10^{-03}$  \\ 
C & 100.0 & 2.0 & 17.8740 & 4.34 $\times 10^7$ & 0.579 & 17.97 & $11625$ & $1.00 \times 10^7$ &  
$5.64 \times 10^{11}$ & 7428 & $4.71 \times 10^{-03}$  \\ 
D & 100.0 & 2.0 & 20.2759 & 2.94 $\times 10^7$ & 0.451 & 16.75 & $10096$ & $2.40 \times 10^7$ &  
$3.22 \times 10^{11}$ & 7613 & $4.17 \times 10^{-03}$  \\ 
E & 100.0 & 2.0 & 17.9534 & 2.95 $\times 10^7$ & 0.507 & 15.82 & $9009 $ & $2.43 \times 10^7$ &  
$4.88 \times 10^{11}$ & 7671 & $4.69 \times 10^{-03}$  \\  
F & 100.0 & 2.0 & 18.4881 & 6.91 $\times 10^7$ & 0.655 & 21.31 & $16342$ & $5.64 \times 10^7$ &  
$4.56 \times 10^{11}$ & 7360 & $4.56 \times 10^{-03}$  \\ 
\hline 
\hline

\end{tabular}
\end{minipage}

\caption[]{The above table contains the simulation name$^a$, the initial redshift$^{b}$,
  the comoving boxsize$^{c}$ [kpc h$^{-1}$],
  the redshift$^{d}$ at the end of the simulation, the total mass$^{e}$ (gas \& dark matter) 
  at the virial radius\footnotemark[2] [$M_{\odot}$], the virial radius$^{f}$ [kpc], 
  the virial velocity$^g$ $(v_{\rm vir}=\sqrt{GM_{\rm vir}/r_{\rm vir}})$ 
  [\kms], the virial temperature$^{h}$ [K], the FoF DM mass$^{i}$ [$M_{\odot}$], 
  the maximum number density$^{j}$ in the halo [$\rm{cm^{-3}}$], the temperature$^{k}$ at the core of 
  the halo [K] and the spatial resolution$^{l}$ [pc] of the simulation. 
  All units are physical units, unless explicitly stated otherwise. }

\label{TableSims}
\end{table*}


\section{The Setup of the numerical simulations} 
\label{Sims} 
\subsection{ The Adaptive-Mesh Refinement Code \enzo}
We have  used the publicly available adaptive mesh refinement
(AMR) code \textsc{Enzo}\footnote{http://enzo-project.org/}. The code has matured 
significantly over the last few years and as of July 2013 is available as 
version \enzolat. Throughout this study we use \enzo version 2.2.
\enzo was originally developed  by Greg  Bryan  and
Mike  Norman at  the University of  Illinois \cite[]{Bryan_1995b, 
Bryan_1997, Norman_1999, OShea_2004, Enzo_2013}. The gravity solver 
in \enzo uses an N-Body particle mesh technique \citep{Efstathiou_1985,  
Hockney_1988} while the hydro calculation are performed using 
the piecewise parabolic method combined with a non-linear Riemann
solver for shock capturing. One of \textsc{Enzo's} greatest strengths lies in the 
fact that additional finer meshes can be laid down as the simulation runs to enhance the resolution
in a given, user defined, region. The  Eulerian AMR  scheme  was  first
developed   by \cite{Berger_1984} and later refined by
\cite{Berger_1989}  to  solve   the  hydrodynamical equations for an
ideal gas. \cite{Bryan_1995b} adopted such a scheme for cosmological simulations.
In addition to this there are also modules available which compute the radiative 
cooling of the gas together with a multispecies chemical reaction network. Numerous
chemistry solvers are available as part of the \enzo package. For our purposes we 
use only the six species model which includes:
${\rm H}, {\rm H}^+, {\rm He}, {\rm He}^+,  {\rm He}^{++}, {\rm e}^-$. In this case 
we are neglecting the effects of \molH cooling - see \S \ref{H2} for more details and limitations 
of this approach.
We also allow the gas to cool radiatively during the course of the simulation. 
Our simulations make extensive use of \textsc{Enzo's} capability to employ nested grids.

\subsection{Nested Grids \& Initial Conditions} 
For our fiducial simulations the maximum refinement level is set to 18.
Our fiducial box size is 2 $\rm{h^{-1}}$ Mpc comoving giving a maximum comoving resolution of 
$\sim  6\ \times 10^{-2}\ \rm{h^{-1}}\ $ pc. At the typical redshifts we are interested in 
($z \sim 15$) this corresponds to a physical resolution of $\sim 4 \times 10^{-3}\ \rm{h^{-1}}\ $ pc.
Furthermore, during the course of the resolution study conducted here (\S \ref{Scaling})
we increase the resolution up to a maximum refinement level of 26. This leads to a maximum 
comoving resolution of $\sim  2\ \times 10^{-4} \rm{h^{-1}}\ $ pc, corresponding to 
$\sim 1\ \rm{h^{-1}}\ AU$ physical at $z \sim 15$.
Initial conditions were generated with the ``inits'' initial 
conditions generator supplied with the \enzo code. 
The nested grids are introduced at the initial conditions stage. 
We have first run exploratory DM only simulations with coarse resolution, 
setting the maximum refinement level to 4. These DM only simulations have a root
grid size of $256^3$ and  no nested grids.
In these exploratory simulations we have  identified the most
massive halo at a redshift of 15 and then  rerun the simulations, 
including the  hydrodynamics module. We also
introduce nested grids at this point. The nested grids  are placed
around the region of interest, as identified from the coarse
DM simulation. We have used four levels of nested grids in our
simulations with a maximum effective resolution of $1024^3$. The
introduction of nested grids is accompanied by a corresponding 
increase in the DM resolution by increasing the number of particles 
in the region of interest. Within the highest resolution region we further restrict the 
refinement region to a comoving region of size $128 \, \rm{h^{-1}}$ kpc around the region 
of interest so as to minimise the computational overhead of our simulations. 
We do this for all of our simulations. 
The  total number of particles in our simulation
is 4935680, with $128^3$ of these  in our highest resolution
region. The grids dimensions at each level at the start of the 
simulations are as follows:
L0[$128^3$],
L1[$64^3$],
L2[$96^3$],
L3[$128^3$].
Table \ref{TableSims} gives further details on the simulations
discussed here. 

\subsection{Comparison to Previous Work}
In comparison to our previous work in this area \citep{Regan_2009} these new simulations
differ in several ways. Firstly, the code used is a newer version of \enzo (version 2.2), 
secondly, the methodology of the simulations is somewhat different. While in \cite{Regan_2009}
we first allowed a large halo to build up before increasing the refinement, we do not do that 
here and instead allow maximum refinement from the start of the simulations and 
follow the gravitational collapse consistently. Thirdly, we increase the resolution of our 
highest resolution simulations significantly - by a factor of 256. Finally, we modified the
refinement criteria and some of the runtime options used by \enzo compared to the 
\cite{Regan_2009} simulations.

\footnotetext[2]{The virial mass is defined as 200 times the mean density of the 
Universe in this case - see \S \ref{virial} for further details.}
\begin{figure*}
  \centering 
  \begin{minipage}{175mm}      \begin{center}
    \centerline{\includegraphics[width=16cm]{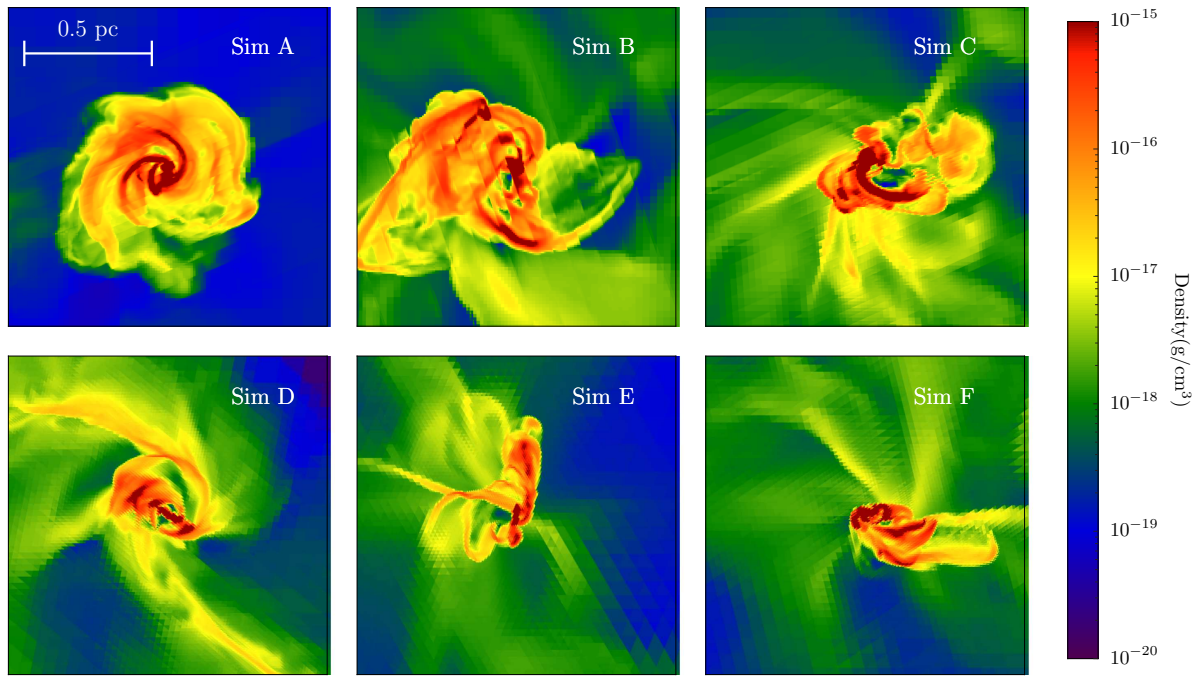}}
    \caption[]
    {\label{MultiPlot}
      Density slices through the centre of the 
      central object in each halo. The ``up'' vector is chosen to be the
      angular momentum vector and so we are looking down onto the central 
      object. Each image is scaled in density and length in the same way. The colour
      scale for the density runs from  $10^{-19}$ g cm$^{-3}$ to $10^{-15}$ g cm$^{-3}$. 
      It is clearly visible that in each case a disk is formed,
      with well defined spiral arms. The size of the plotted region in each panel 
      is 1pc. }
   \end{center} \end{minipage}
\end{figure*}


\subsection{Refinement Criteria}
\label{Collapse} 

\enzo uses adaptive grids to
provide increased resolution where it is required. 
For the simulations discussed in this paper we have used
three refinement criteria implemented in \enzo: DM over-density, baryon over-density, 
and Jeans length. The first two criteria introduce 
additional meshes when the over-density (${\Delta \rho \over \rho_{\rm{mean}}}$) of a 
grid cell with respect to the mean density exceeds 3.0 for baryons and/or DM. The
third criterion has received quite a lot of attention in recent 
years with several groups arguing for a higher threshold, than the canonical
Truelove criterion of 4 cells \citep{Truelove_1997}. \cite{Federrath_2011}, 
\cite{Turk_2012} and \cite{Latif_2013a} have all shown that simulations require a minimum Jeans 
length resolution of 32 cells in order to obtain converged turbulent energy results. 
We follow \cite{Latif_2013a} and \cite{Meece_2013} in using a Jeans
resolution criteria of 64 cells in order to fully resolve the turbulent 
energy requirement in the highest density regimes. 
We set the \emph{MinimumMassForRefinementExponent} parameter to $-0.1$ making the
simulation super-Lagrangian and therefore reducing the threshold for
refinement as higher densities are reached \cite[]{OShea_2008}. We 
furthermore set the \emph{MinimumPressureSupportParameter} equal to 
ten as we have restricted the maximum refinement level in our 
simulations \cite[e.g.][]{Kuhlen_2005}. When this option is selected the code defines a minimum
temperature for each grid at the highest refinement level. This minimum temperature is that 
required to make each grid Jeans stable multiplied by the above parameter. This parameter
was introduced into \enzo to alleviate artificial fragmentation and angular momentum 
non-conservation - see \cite{Machacek_2003} for further details.
We further make use of the 
\emph{RefineRegionAutoAdjust} parameter setting it to one. This parameter
modifies the refinement region during the course of a simulation so as to allow
only the highest resolved dark matter particles into the refinement region. 
Larger mass dark matter particles are therefore excluded. We then allow our simulations 
to evolve restricted only by the maximum refinement threshold.

\subsection{Zero Metallicity and H$_{2}$ Dissociation}
\label{H2}

Assuming zero metallicity and efficient dissociation of $H_2$ \citep{Wise_2008, 
Regan_2009, Johnson_2011,Prieto_2013, Latif_2013c, Latif_2013a, Latif_2013b, Latif_2013d}
strongly reduces fragmentation and therefore optimises conditions for direct collapse 
of gas into a massive (seed) black hole.  As discussed extensively in the literature 
these conditions may  be fulfilled in a small fraction of haloes above the atomic cooling threshold 
which may nevertheless be sufficiently abundant to act as seeds for the most massive supermassive 
black holes at high redshift \citep[e.g.][]{Loeb_1994, Bromm_2003, Spaans_2006, Dijkstra_2008}. 
Accurate modelling of the metal enrichment history of these haloes is very challenging. 
\cite{Cen_2008} have argued that the mixing of metals produced by early episodes of 
population III (POP III) star formation is very inefficient, but as pointed out by 
\cite{Omukai_2008} even small amounts of metal contamination may be enough to induce 
fragmentation. Efficient dissociation is required to allow the build up of a halo to 
$T_{\rm vir} > 10000\ {\rm K} $ without collapse. In order to efficiently 
dissociate $H_2$  a strong UV background is required. Recent work by several groups 
\citep{Shang_2010, Wolcott-Green_2011, Johnson_2013} has determined that a Lyman-Werner 
flux of $\sim$  1000 J$_{21}$ for a UV background with a POP III 3 spectrum or $\sim$ 
30 - 300 J$_{21}$ for a POP II spectrum is required to allow for the build up of an atomic 
cooling halo, where $J_{21}$ is the canonical background intensity in units of 10$^{-21}$ erg cm$^{-2}$
s$^{-1}$ Hz$^{-1}$ sr$^{-1}$. Here we also make the assumption that in our simulations both metal and 
$H_2$ cooling are inefficient.  Such haloes will be rare, but may exist when a local strong UV 
source exists $\lesssim 10$ kpc from the collapsing halo which augments the global UV background 
in the early Universe \citep{Dijkstra_2008, Agarwal_2012}.


\section{Results of the Numerical Simulation}
\label{results}

\subsection{The Suite of Simulations}
We have performed 6 simulations of haloes with virial temperatures 
in the range of 9000 K to 16000 K and virial velocities in the range of 
15 to 22 km/s collapsing at redshifts $15 < z < 22$ all within a 2 Mpc
h$^{-1}$ box starting at $z = 100$.  The details of each simulation
are  shown in Table \ref{TableSims}. Each simulation was run 
to a maximum refinement level of 18 with an initial root grid of dimension 128 
grid cells. This results in a maximum comoving resolution of $5.96 \times 10^{-2} 
\rm{h^{-1}}$ pc. The physical resolution achieved in each run is given in Table 
\ref{TableSims}. Once the simulations reach the maximum refinement level the collapse is able 
to continue adiabatically. Given the short dynamical times at the maximum refinement 
level the evolution of all, but the central regions has effectively stopped. As gas continues
to fall onto the central regions, thereby increasing the densities, the cooling times
of the gas continue to decrease. As the maximum refinement level has already been reached 
the code is no longer able to calculate accurately the gas properties and errors in the 
hydro solver begin to appear. At this point we terminate the simulation, discarding any 
spurious outputs. In addition to the above simulations with a maximum refinement level
set to 18 we further run simulations A and C with 4 different maximum refinement settings 
(see Table \ref{ScalingTable} for more details). We are thus able to see the effect of 
numerical resolution on the central object obtained. We discuss this further in \S \ref{Scaling}.
Note that unlike in our previous simulations \citep{Regan_2009}, where we only 
started refinement when a simulated halo was well above the atomic cooling threshold, 
we have followed here the growth of the virial temperature of the six individual haloes 
across the atomic cooling threshold with the code always allowed to refine where necessary.

\begin{figure*}
  \centering 
  \begin{minipage}{175mm}      \begin{center}
    \centerline{\psfig{file=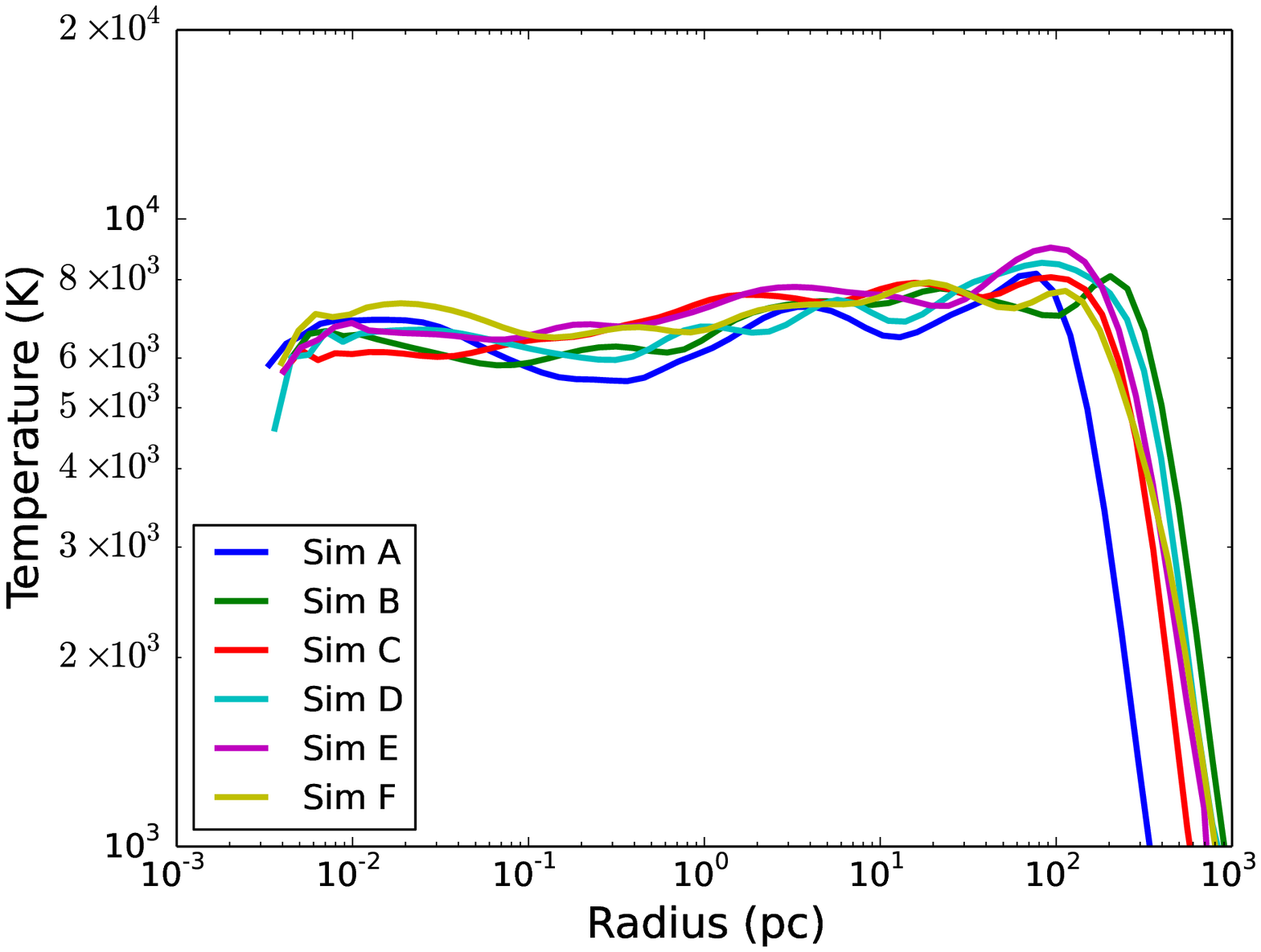,width=9cm}
      \psfig{file=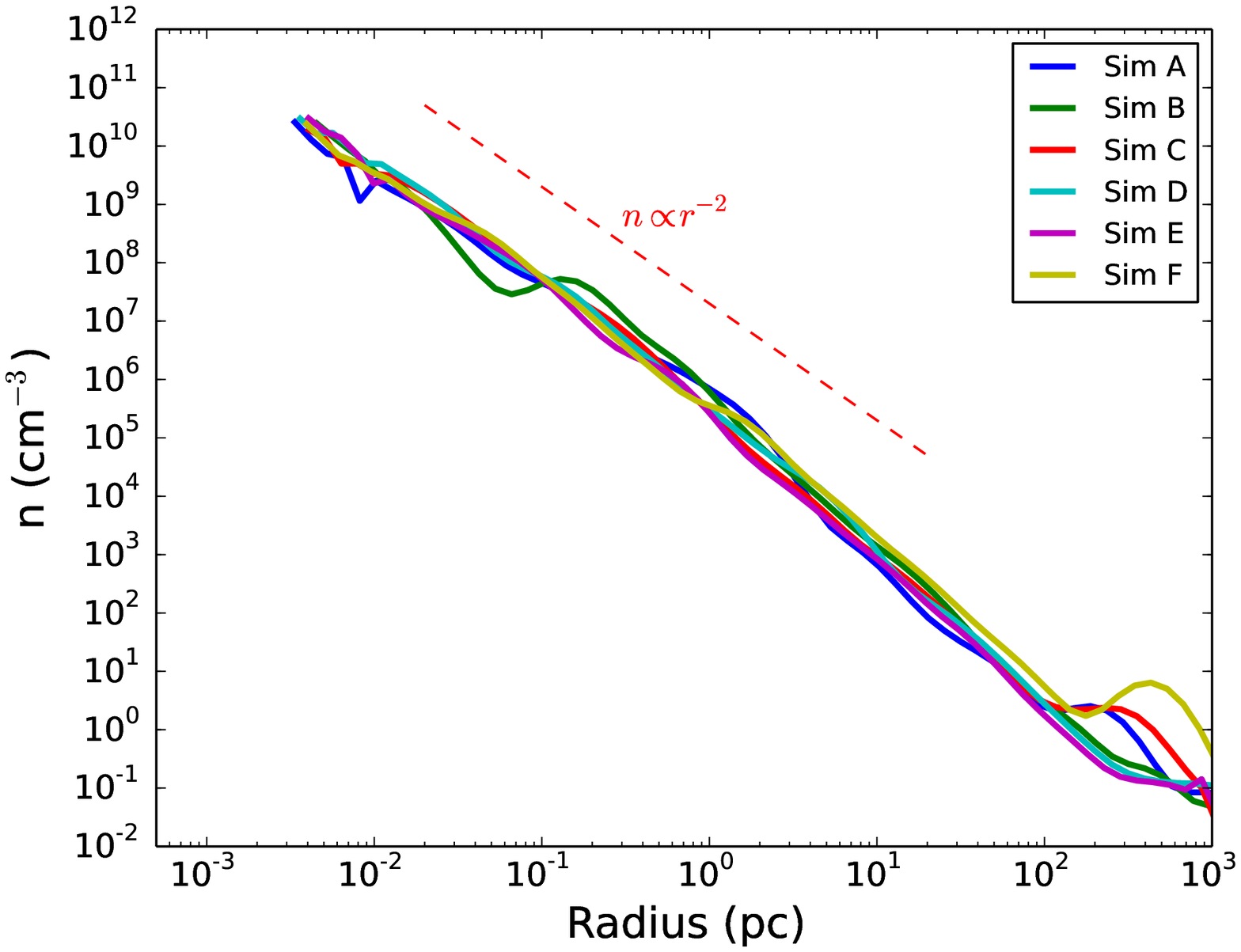,width=9cm}}     
    \caption[]{\label{DensityProfile}
      {\it Left Panel:} 
      The temperature profile for the gas within the halo. As expected for a 
      collapse mediated by atomic hydrogen cooling the temperature remains constant 
      at approximately $T\sim 6000 - 8000 \rm \ K$ during the collapse. 
      {\it Right Panel:} 
      The density profile for each halo. Number densities reach values of up to 
      $n\sim 1 \times 10^{11} \ \mathrm{cm^{-3}}$ with a mean slope of $\rho \propto r^{-2.02}$ 
      (i.e. an isothermal profile)}
   \end{center} \end{minipage}
\end{figure*}
 

\subsection{Simulation Morphologies} \label{Morph}

In Figure \ref{MultiPlot} we show visualisations of the central pc 
of our simulations containing about $10^4$ \msolar at a time when the
gas is settling  into (marginal) rotational support. In all cases
unstable discs have formed which develop prominent spiral features. 
As in our previous simulations and that of \cite{Latif_2013a, Latif_2013c,
Latif_2013d} $10^4$ \msolar appears to be the characteristic mass scale for which 
the collapse becomes marginally rotationally supported and angular momentum support 
first sets in. Note, however, that as we will discuss later the innermost  part of the 
halo nevertheless continues to collapse. The panels are density slices taken through a 
plane perpendicular to the angular momentum vector, calculated at the core of the halo. 
We are therefore looking down onto the central plane of the disc/spiral object that forms. 
The figure clearly shows the formation of a well defined central object. Simulations A, B 
\& D show very well defined spiral arms while simulations C, E \& F show a more complex
morphology. As we will see in more detail later, formation of a (marginally) dynamically 
stable and rotationally supported object is the generic outcome of our simulations.  \\

\subsection{Properties of the Gas in the Dark Matter Haloes when Collapse Occurs} 
\label{virial}

The collapse of the gas in our simulations starts when the virial
temperature of the halo has reached the atomic cooling threshold.  
The virial temperatures of our haloes are therefore all in the
narrow range of 9000 K - 16000 K with an average virial temperature at which 
collapse begins of $\rm{T_{vir}} \sim 11000$ K. The virial radius of the collapsed  
object is approximately 0.5 kpc in all simulations. The virial quantities are defined 
such that the density at the virial radius is 200 times the mean density of the Universe 
at that redshift. For the haloes found here this corresponds to a virial mass of a few times
$10^7$ \msolar \citep{Mo_2002}. That all of our haloes have 
similar masses is due to the maximum refinement criteria we have imposed. \\ \indent
Once collapse begins, additional refinement is engaged by the code. This extra refinement 
allows us to track the collapse down to sub parsec scales, but effectively freezes any further
growth of the  halo. Hence, in a simulation such as this, with a relatively high maximum 
refinement level, following the growth of haloes to much higher masses is currently extremely
challenging.  We will return to this issue in \S \ref{Scaling} where we conduct further 
simulations with different maximum resolutions.  

\begin{figure*}
  \centering 
  \begin{minipage}{175mm}      \begin{center}
    \centerline{
      \psfig{file=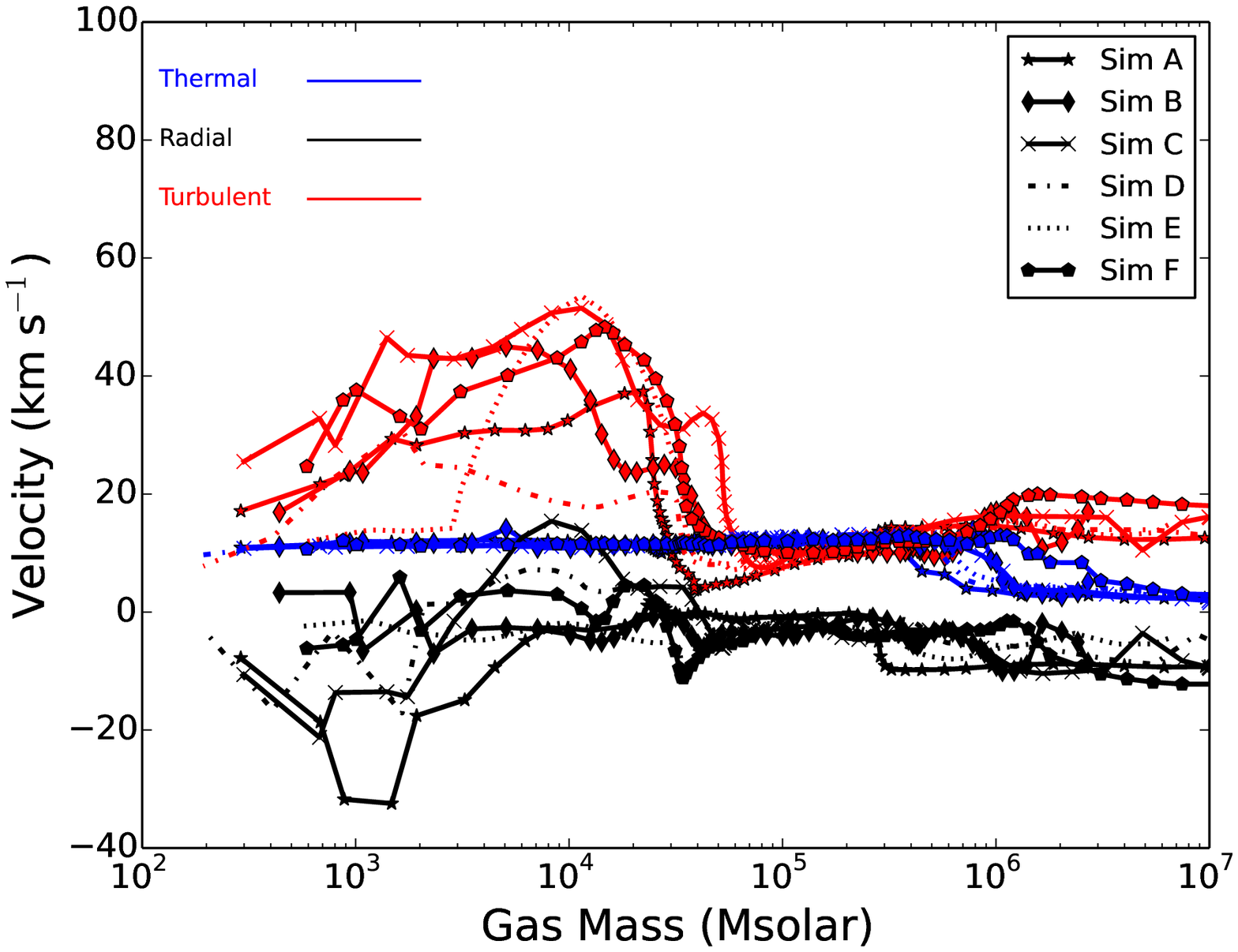,width=9cm}
      \psfig{file=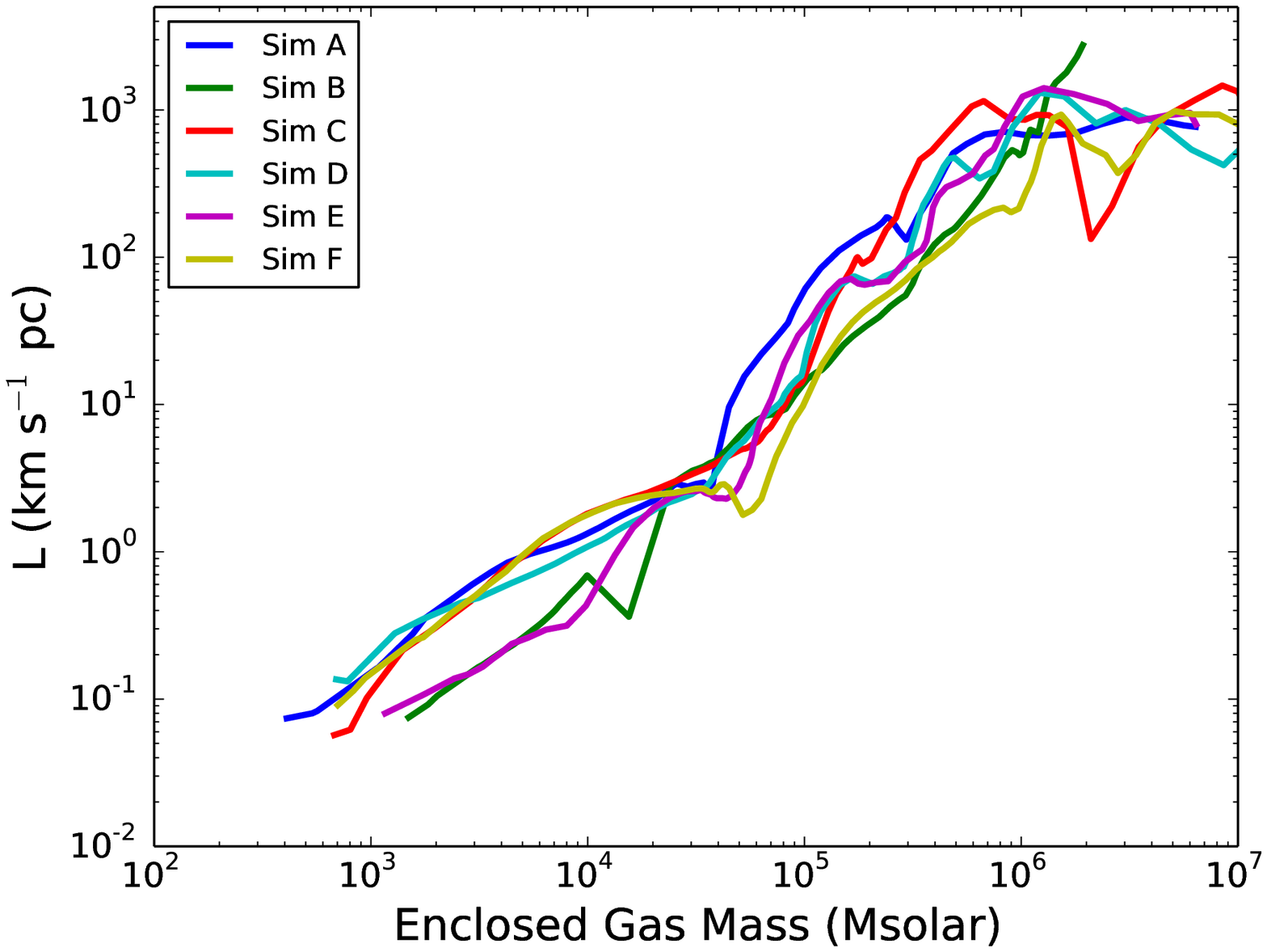,width=9cm}
    }
    \caption[]{\label{VelocityEvolution}
      {\it Left Panel:} 
      The velocity profile for three distinct velocities. In blue we have plotted 
      the thermal velocity profile. In red the turbulent velocity profile and 
      finally in black the radial velocity profile of the gas. 
      {\it Right Panel:} 
      The central object loses angular momentum allowing the gas to fall 
      to the centre. In \cite{Regan_2009} we showed the evolution of a central 
      object and found that it can lose $\approx$ 90\% of its initial angular
      momentum. }
   \end{center} \end{minipage}
\end{figure*}
 

\begin{figure*}
  \centering 
  \begin{minipage}{175mm}      \begin{center}
    \centerline{
      \psfig{file=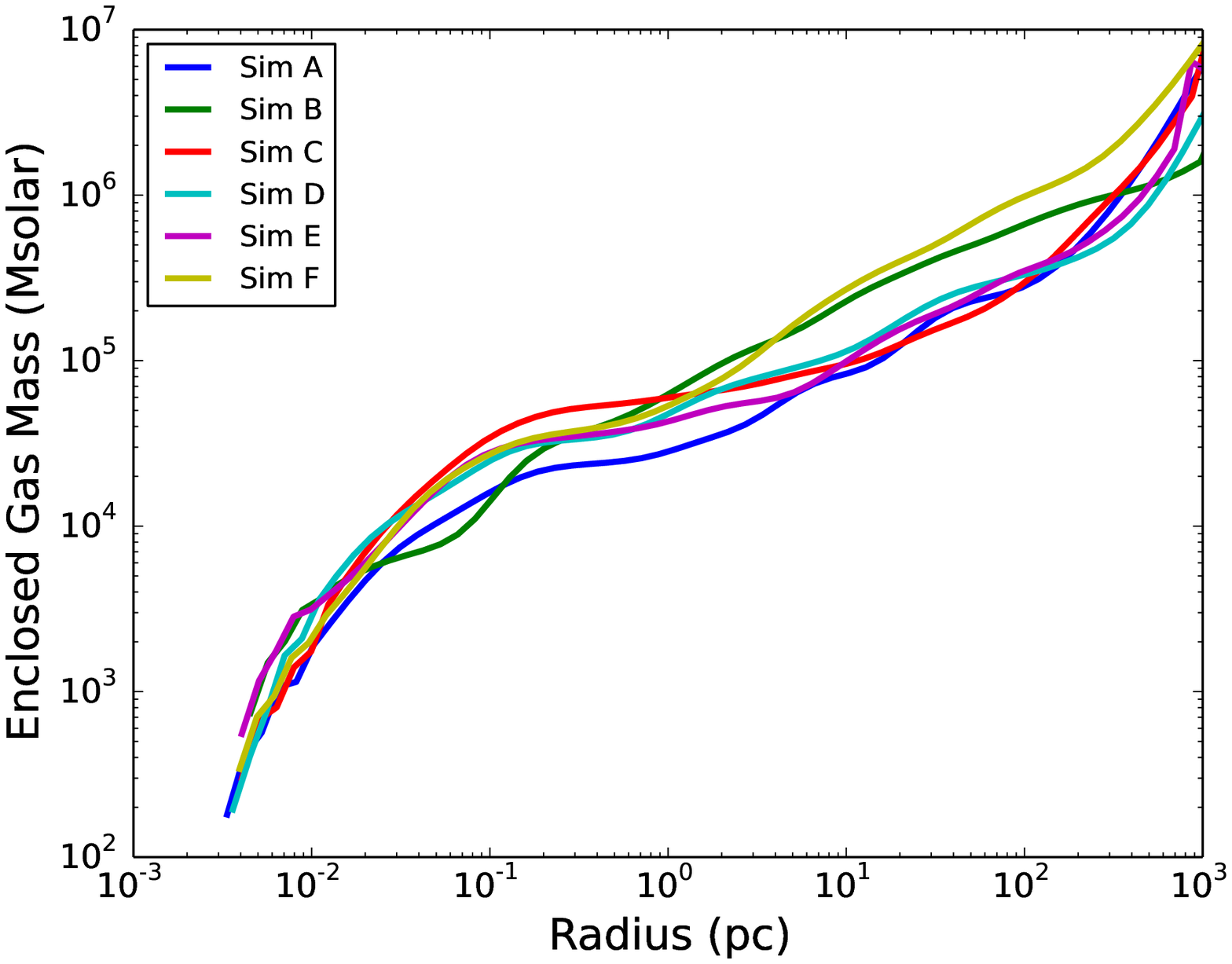,width=9cm}
      \psfig{file=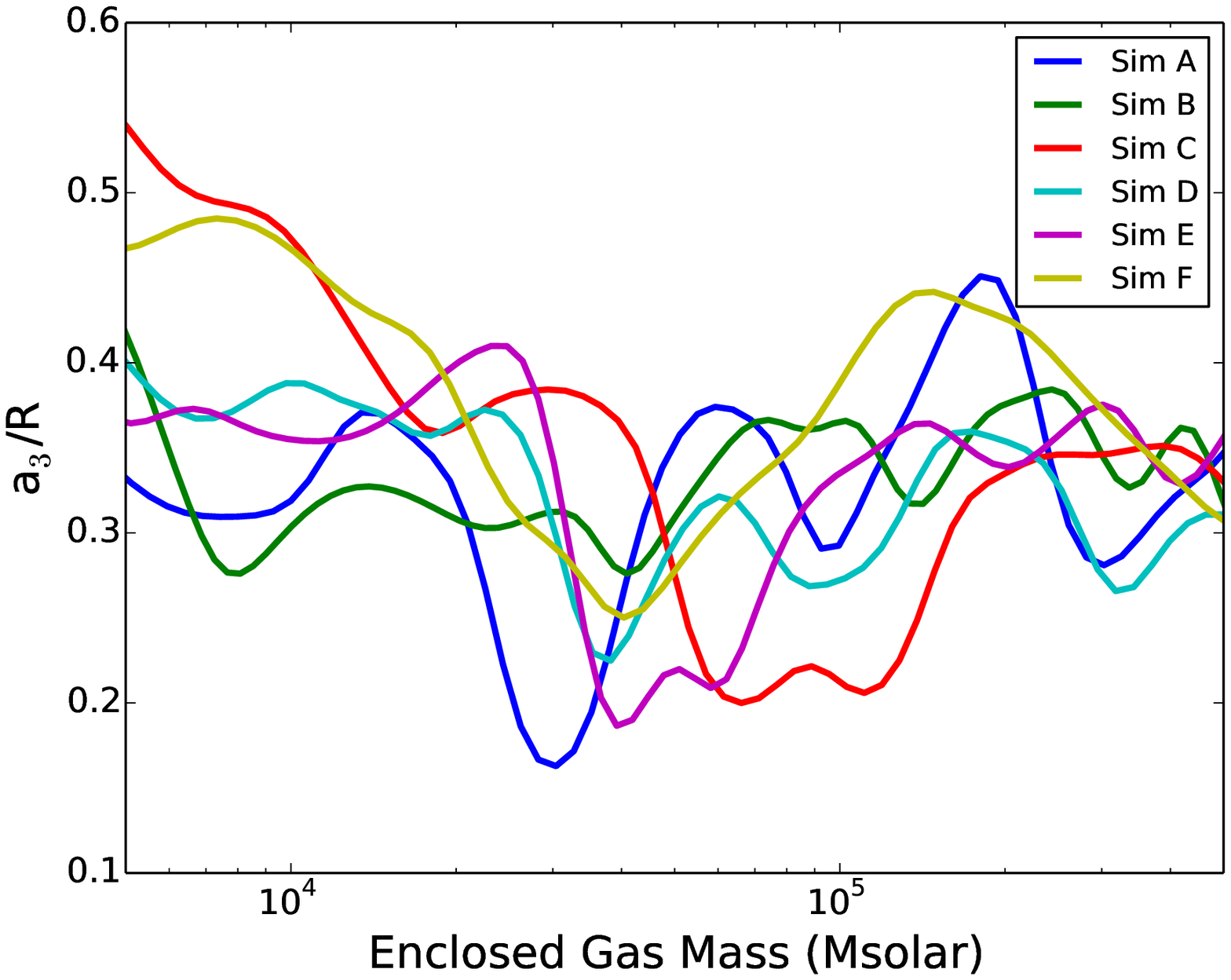,width=9cm}}     
    \caption[]{\label{EnclosedMass}
      {\it Left Panel:} 
      The enclosed mass plotted against radius. The curve flattens as 
      mass settles into rotational support at a given radius. The settling corresponds 
      to the formation of a disk. In the simulations a disk forms between 
      0.1 and 1 pc. 
      {\it Right Panel:}
      The ratio of the minimum eigenvalue found from the inertia tensor and
      the radius of the central object at a given point plotted against enclosed 
      mass. Dips in the ratio indicate that a disk is forming. }
   \end{center} \end{minipage}
\end{figure*}
 

\begin{figure*}
  \centering 
  \begin{minipage}{175mm}      \begin{center}
    \centerline{
      \psfig{file=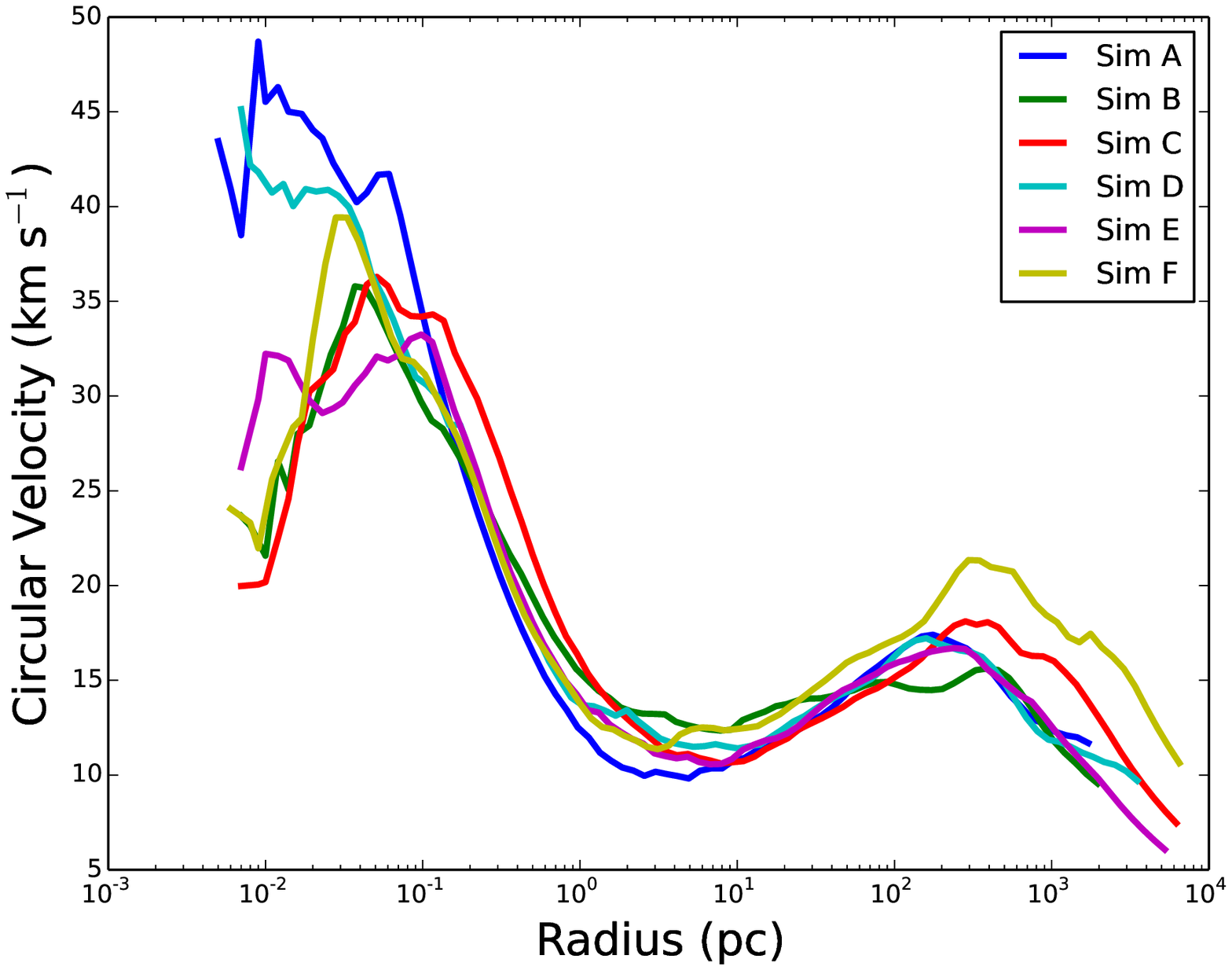,width=9cm}
      \psfig{file=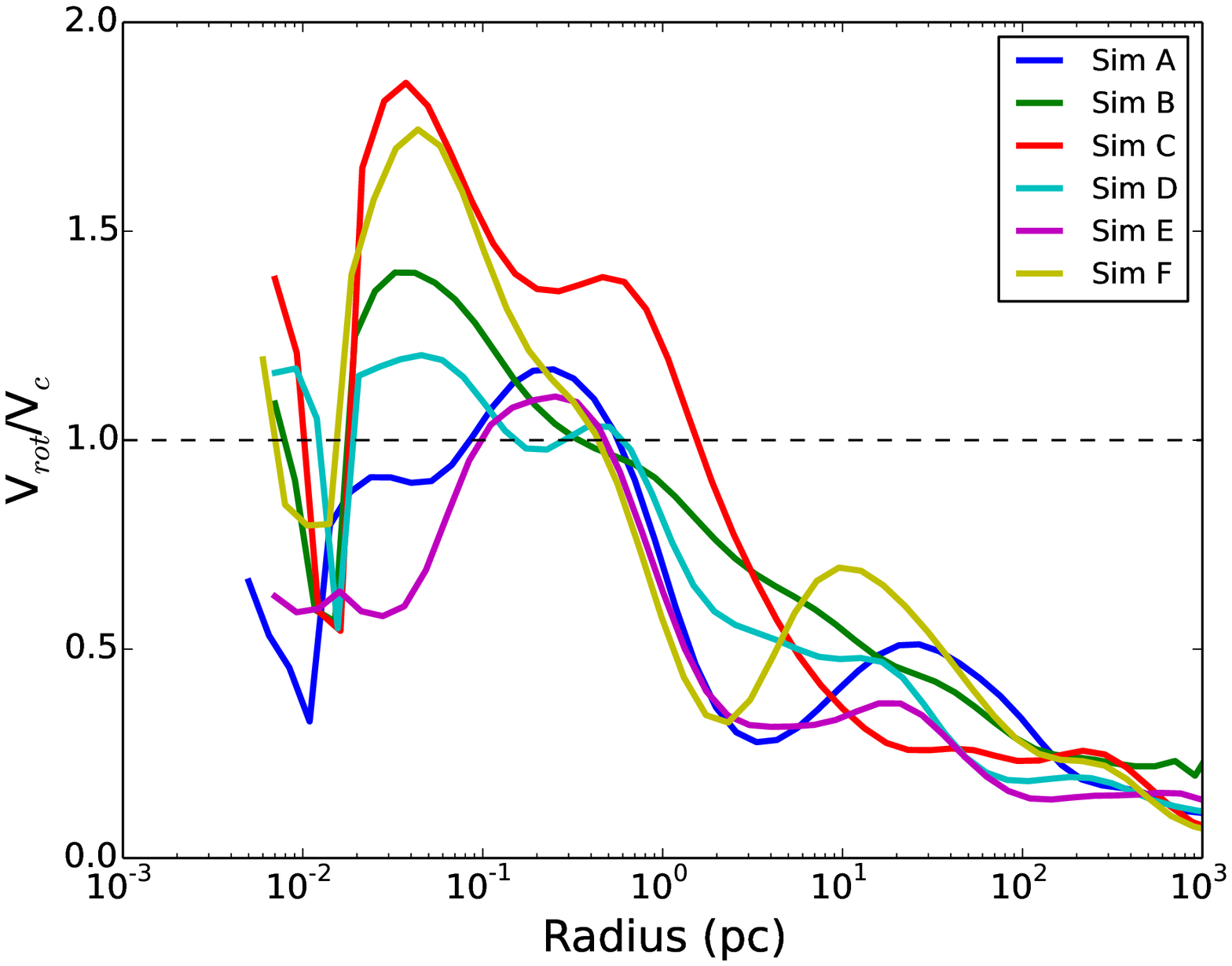,width=9cm}}     
    \caption[]{\label{RotSupport}
      {\it Left Panel:} 
      The circular velocity profile for our six fiducial simulations.
      {\it Right Panel:}
      The ratio of rotational velocity to circular velocity plotted against 
      radius. Each simulation achieves a value close to unity at radii between 
      0.1 and 1.2 pc.}
   \end{center} \end{minipage}
\end{figure*}

\subsection{Profiling the Central Object}

As cooling is facilitated only via atomic hydrogen cooling
in the simulations performed here, the  gas
cannot cool below approximately 7000 K. In Figure \ref{DensityProfile} (left panel) we 
show the temperature of the gas over several decades in radius. We
plot the temperature out to
approximately the virial radius which is well outside the realm of the collapse. 
The temperature of the gas is found by averaging the temperature of cells in spherical 
shells outwards from the densest point in the halo. As expected, the temperature remains
approximately constant as the density grows towards the centre of the halo. Initially the gas
is shock-heated to $T \sim 10^4$ K, close to the virial radius, from where it cools via atomic
hydrogen transitions to $T \sim 7000$ K. \\ \indent 
Figure \ref{DensityProfile} (right panel) shows the gas density profile of the halo 
over the same range. The density profile is initially quite flat at, or outside, the 
virial radius, but quickly steepens to attain a slope of $n \propto r^{-2}$ as 
expected for an isothermal collapse. The profile is not completely smooth due to the 
presence of small dense clumps within the halo (as shown in Figure \ref{MultiPlot}). 
All simulations show similar profiles with the maximum number density obtained in 
each simulation found to be $\approx 1 \times 10^{11}$ cm$^{-3}$.  The result here are
in very good agreement with those of our own previous simulations as well
as those of other authors \citep{Wise_2008, Latif_2013a, Latif_2013c, Latif_2013d}.
The maximum density reached is a function of the maximum resolution of the simulation. 
As we will see in \S \ref{Scaling} higher resolution simulation are able to follow the
collapse to higher densities at the expense of failing to 
track the further evolution of the gas at lower resolution (lower density). \\ \indent 
In the left hand panel of Figure \ref{VelocityEvolution} we have plotted the thermal 
velocity (blue lines), turbulent velocity (red lines) and the radial velocities (black) 
lines against enclosed mass. Again we plot out to the virial mass. The thermal velocity 
is computed as $V_{TH} = \sqrt{3 k_B T/M}$, where  $k_B$ is the Boltzmann constant, 
$T$ is the temperature of the gas in a given shell and $M$ is the gas mass in that shell. 
The turbulent velocities are calculated by computing the root mean square velocity of the 
gas after subtracting the centre of mass velocity of the halo and the velocity due to the 
radial inflow of the gas. Finally, the radial velocity of the gas is determined by computing the 
radial component of the Cartesian velocities in each shell. \\ \indent 
Near the virial radius the turbulent velocities are stable at approximately 20 \kms and 
similar to the virial velocity of the DM halo, the thermal velocity is very low and radial 
in-fall is constant at approximately 15 \kms. As the gas falls towards the centre the 
turbulent velocities grow and exceed the thermal and radial components. All of the simulations 
show significant turbulent velocities consistent with their complex morphology as seen in 
Figure \ref{MultiPlot}. The radial velocities have a mean value of close to 10 \kms, for the 
majority of outputs. Simulations C, D and F have radial velocities which display a significant 
positive contribution at an enclosed mass of about $10^4$ \msolarc. At this point the central 
object becomes rotationally supported and the radial velocity switches sign. Significant 
amounts of gas are flung out from the central core along the spiral arms resulting in 
average positive values of the radial velocity. The same is seen -- albeit to a
lesser extent --  in each of the other simulations. \\ \indent
In the right hand panel of Figure \ref{VelocityEvolution} we plot the angular momentum 
of the central object against the enclosed gas mass in spherical shells. The angular momentum 
is calculated in spheres centred on the densest point in the simulation. As discussed in 
detail in \cite{Regan_2009} the innermost gas shells ``lose'' up to 90\% of their initial 
angular momentum during the turbulent collapse probably due to a combination of angular momentum 
redistribution and angular momentum cancellation. The low angular momentum gas that ends up 
flowing to the centre then  settles into a rotationally supported object. We determine the 
rotational properties of the collapsing gas in a similar way as in \cite{Regan_2009}.  We 
calculate the inertia tensor $\tilde{I}$ and calculate  its eigenvectors to describe the 
principal axes of the rotation of the central object.


The angular momentum and the inertia tensor are related as, 
\begin{equation} 
\vec{l} = \tilde{I} \vec{\omega},
\end{equation}

\noindent where $\vec{\omega}$ is the angular  velocity. Using the square root of
the largest eigenvalue of the inertia tensor, $a_{1}$, we  
then estimate the rotation velocity as 

\begin{equation} \label{RotVelocity}
V_{\rm rot} \approx  { |\overrightarrow{l}| \over a_{1}}.
\end{equation}

\noindent  As in \cite{Regan_2009} we then use the square root of the
smallest eigenvalue of the inertia tensor as a proxy for the thickness of 
the flattened object  formed. \\ \indent 
In the left panel of Figure \ref{EnclosedMass} we show the enclosed gas mass against 
radius for the six simulations. In each simulation the enclosed mass 
first decreases linearly inwards as expected for an isothermal density
profile. The gas at these radii is virtually in free-fall, collapsing
turbulently at a significant fraction ($\approx 25-30 \%$) of the free-fall speed. 
Between 1 and 0.1 pc the curve of enclosed mass flattens indicating 
that the collapse becomes marginally rotationally supported. The onset of rotational support  
is also reflected in the right panel of Figure \ref{EnclosedMass} where we plot the ratio of the 
square root of the smallest eigenvalue of the inertia tensor, a3, and the radius at that point. 
Note that the plot focuses on the range of mass shells where the rotational support sets in between 
$5 \times 10^3$\msolar \ and $5 \times 10^5$\msolar. All of the simulations show a 
characteristic dip in ``thickness''   between $\sim 1 \times 10^4 $ \msolar and 
$\sim 1 \times 10^5 $ \msolar. The dip  is due the formation of a 
flattened disk-like structure. Interestingly simulations A, C \& E show the most 
prominent dip at a few times $ 10^4 $ \msolar. This corresponds to a radius of $\sim 1$ pc. 
Figure \ref{MultiPlot} clearly shows a flattened object with spiral arms for simulation A, but 
simulations C \& E have a more complex morphology. Nonetheless the calculations indicate 
the presence of a fat disk (since the ratio of a3/R is relatively large). 
Simulation B shows no prominent dip compared to the surrounding gas, rather the 
ratio is constant at $\sim 0.3$. Simulation D shows a double dip at 
enclosed masses of $\sim 3 \times 10^4 $ \msolar and $\sim 1 \times
10^5 $ \msolar consistent with the visual impression from Figure  \ref{MultiPlot}. 
Simulation F shows a wide dip at $\sim 4 \times 10^4 $ \msolar which again is 
well matched by the  visual impression from Figure \ref{MultiPlot}.

\begin{figure*}
  \centering 
  \begin{minipage}{175mm}      \begin{center}
    \centerline{
      \psfig{file=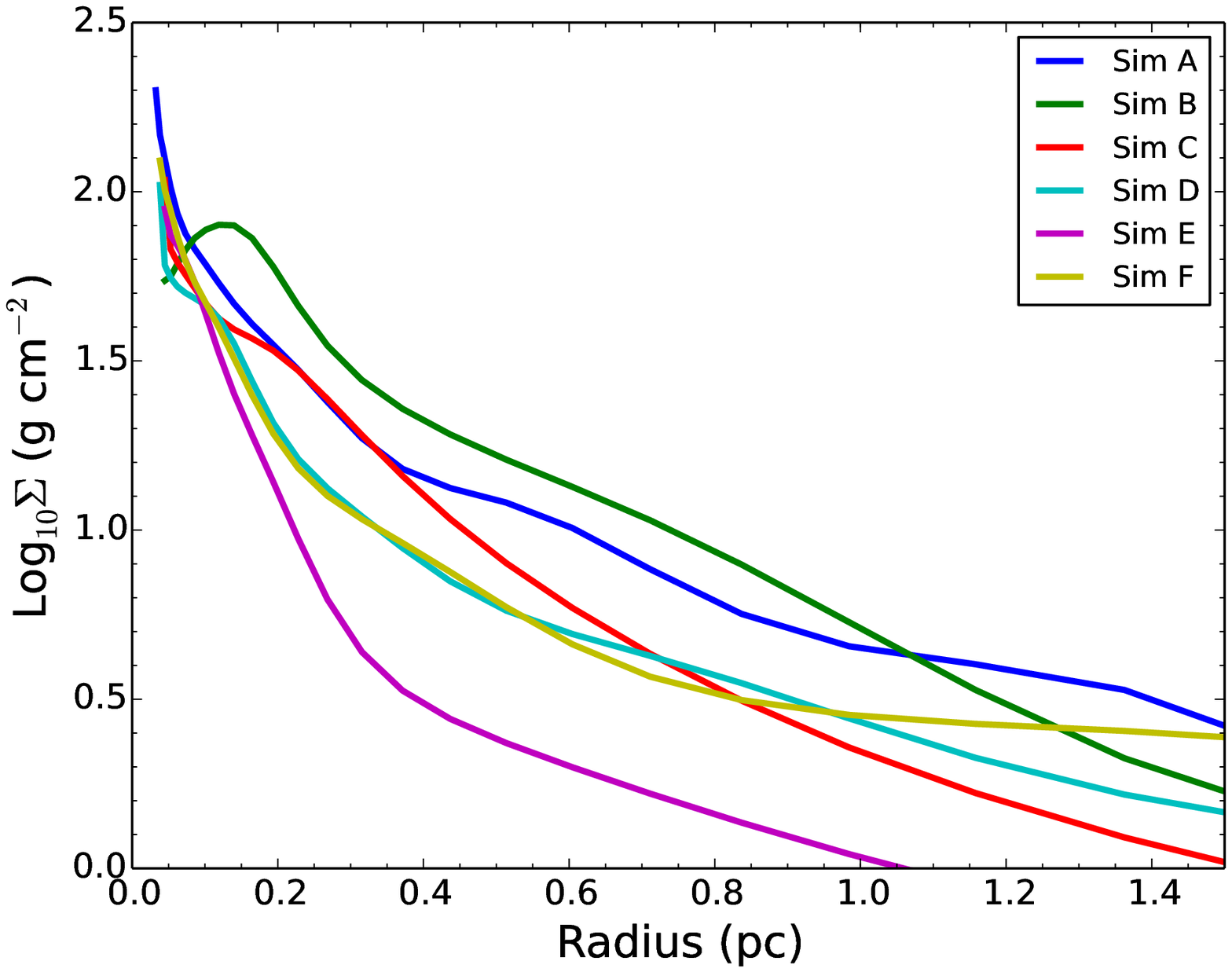,width=9cm}
      \psfig{file=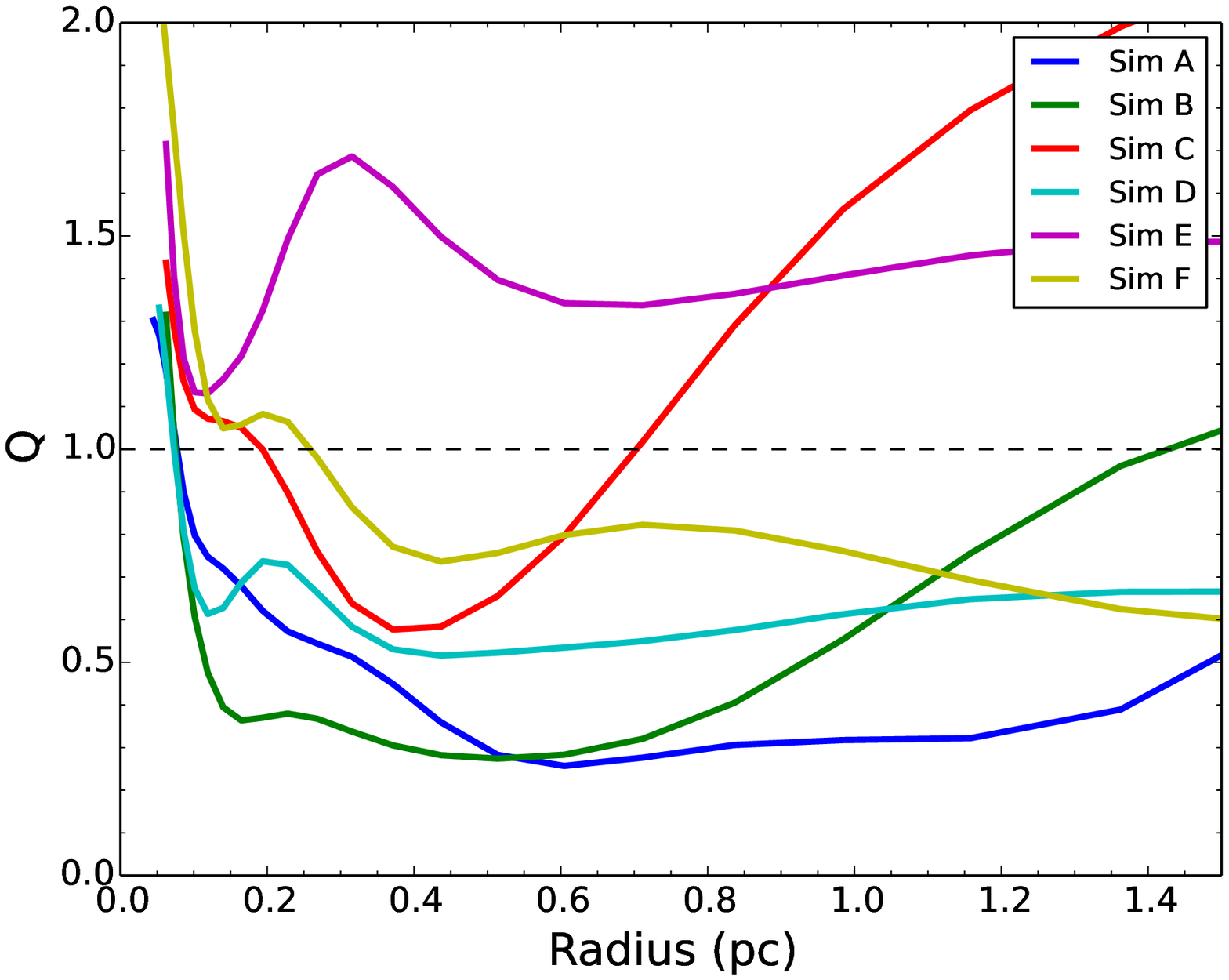,width=9cm}}
    \caption[]{\label{Toomre}
      {\it Left Panel:} 
      The surface mass density calculated in the plane of the disk. Each disk 
      displays an exponential profile. The average scale length of the disks 
      formed is $\sim 0.41$ pc.
      {\it Right Panel:}
      The Toomre parameter plotted against radius. The Toomre parameter was 
      calculated in the plane of the disk.}
  \end{center} \end{minipage}
\end{figure*}

\subsection{Rotational Support and Gravitational Stability}

We now investigate the actual level of rotational support in the collapsing gas in our 
simulations by comparing the rotational velocity to the circular velocity, 
\begin{equation}
V_{\rm cir} =  \sqrt{\frac{GM(R)}{R}}
\end{equation}
In the left  panel of Figure \ref{RotSupport} we show
the circular velocity for our six haloes and in the right panel we
show the ratio of rotational and circular velocity  plotted against
radius. The rotational velocity is computed using Eq. \ref{RotVelocity}. 
Between $\sim 1$ pc and  $\sim 1 \times 10^{-2}$ pc  the ratio 
of rotational to circular velocity reaches  a value greater than 1.0  
in all simulations - right panel of Figure \ref{RotSupport}. 
To investigate the stability of the disks formed we also calculate their
surface mass density and the corresponding Toomre parameter as a function of radius.  
Similar to \cite{Regan_2009} we find  exponential surface mass density profiles 
(left panel of Figure \ref{Toomre}) with scale lengths in the range 
$\rm{R_d} \sim 0.24 - 0.53$ pc. The Toomre stability parameter shown in the right 
panel of Figure \ref{Toomre} is calculated as \cite[]{Toomre_1964},
\begin{equation}
Q (r) = {c_{\rm{s}} \kappa \over \pi G \Sigma},
\end{equation}  
where $c_{\rm s}$ is the sound speed, $\kappa = \sqrt{2} \, (V_{\rm{rot}}
/ r) \, (1 + \rm{d \, ln \, V_{\rm{rot}}} / \rm{d \, ln \, r})^{1/2} $ 
\cite[]{Oh_2002, Binney_2008} is the epicyclic frequency, $V_{\rm{rot}}$ is the 
rotational velocity, $\Sigma$ is the surface mass density and $r$ is the radius. 
For values of $Q < 1$ the disc is expected to be gravitationally unstable.  
Simulations C, E \& F all have  values $Q >$ 1 within a scale length or more. 
Simulations A, B \& D are approaching gravitational stability, but do not 
have values $>$ 1.0 at the end of the simulation. 
These results should not be surprising given the rather unrelaxed dynamical 
state of the discs as they undergo further collapse and evolution. Overall the 
discs are found to be marginally stable with some fragmentation and formation of clumps 
during the course of the collapse. Small scale turbulence
effects within the disk, where $\rm{Q \lesssim 1}$, may furthermore induce fragmentation 
\citep[e.g.][]{Hopkins_2012, VanBorm_2013}. \cite{Latif_2013a, Latif_2013b} have recently employed
a subgrid turbulence model to investigate the effect of turbulence at small scales. They
found that while the general properties of a collapse are unchanged with the subgrid model, 
fragmentation is increased slightly and the formation of discrete clumps is more pronounced. 
We will discuss the formation of bound clumps further in \S \ref{Clumps}.
The morphology of the marginally rotationally supported gas varies rapidly
between different simulations, but the formation of a central rotationally 
supported (marginally) gravitationally stable disc appears to be a robust 
outcome of our simulations. 

\subsection{Effects of Environment on Halo Angular Momentum Dissipation}
As noted in the introduction we conducted a study using our six fiducial simulations 
on the environmental dependence of haloes on early angular momentum dissipation. 
We investigated the dependence of the early dissipation of gas angular momentum on 
halo ``rareness'' which we quantified using the peak height relative to the RMS fluctuation 
of the density field of each halo. As discussed in more detail in Appendix \ref{appendixA}
we could not identify any significant trend with the rareness of the halo.


\begin{table*}
\centering
\begin{minipage}{160mm}
\begin{tabular}{ | l | l | l | l | l | l | l | l | l | l}
\hline \hline 
\em{ Sim$^a$ }
& \textbf{\em $z_{end}$$^{b}$} & \textbf{\em $M_{200}$$^{c}$}  & \textbf{\em $R_{200}$$^{d}$}
& \textbf{\em $V_{200}$$^{e}$} & \textbf{\em $T_{vir}$$^{f}$}  & \textbf{\em $M_{DM}$$^{g}$} 
& \textbf{\em $\rho_{max}$$^{h}$} & \textbf{\em $P_{DM}$$^{i}$} & \em {$\Delta$ R$^{j}$} \\ 
\hline 
A12 &  20.15 & 3.70 $\times 10^7$ & 0.490 & 18.04 &  $11709 $ & $3.02 \times 10^7$&  
$5.33 \times 10^{7}$ & $8.301 \times 10^2$ & $2.67 \times 10^{-01}$  \\
A14 & 20.80 & 3.50 $\times 10^7$ & 0.467 & 17.96 & $11609 $ & $2.87 \times 10^7$ &  
$6.40 \times 10^{8}$ & $8.301 \times 10^2$ & $6.53 \times 10^{-02}$  \\ 
A18 & 21.89 & 2.58 $\times 10^7$ & 0.402 & 16.63 & $9961$ & $2.10 \times 10^7$ &  
$3.55 \times 10^{11}$ & $8.301 \times 10^2$ & $3.89 \times 10^{-03}$  \\ 
A22 & 21.90 & 2.57 $\times 10^7$ & 0.401 & 16.61 & $9935$ & $2.10 \times 10^7$ &  
$4.44 \times 10^{12}$ & $6.386 \times 10^1$ & $2.43 \times 10^{-04}$  \\ 
A26 & 21.90 & 2.58 $\times 10^7$ & 0.401 & 16.62 & $9948 $ & $2.10 \times 10^7$ &  
$3.66 \times 10^{11}$ & $6.386 \times 10^1$ & $1.52 \times 10^{-05}$  \\  

C12 & 16.50 & 9.22 $\times 10^7$ & 0.803 & 22.23 & $17790$ & $7.52 \times 10^7$  &  
$7.52 \times 10^{7}$ & $8.301 \times 10^2$   & $3.25 \times 10^{-01}$  \\ 
C14 & 17.50 & 5.14 $\times 10^7$ & 0.625 & 18.81 & $12732$ & $4.18 \times 10^7$&  
$2.78 \times 10^{8}$ & $8.301 \times 10^2$  & $7.67 \times 10^{-02}$  \\
C18 & 17.87 & 4.35 $\times 10^7$ & 0.579 & 17.97 & $11625 $ & $3.54 \times 10^7$ &  
$5.64 \times 10^{11}$ & $8.301 \times 10^2$ & $4.71 \times 10^{-03}$  \\ 
C22 & 17.88 & 4.32 $\times 10^7$ & 0.578 & 17.94 & $11588 $ & $3.53 \times 10^7$ &  
$1.03 \times 10^{14}$ & $6.386 \times 10^1$ & $2.94 \times 10^{-04}$  \\ 
C26 & 17.85 & 4.39 $\times 10^7$ & 0.582 & 18.02 & $11693$ & $3.60 \times 10^7$ &  
$2.90 \times 10^{16}$ & $6.386 \times 10^1$ & $1.84 \times 10^{-05}$  \\ 

\hline 
\hline

\end{tabular}
\end{minipage}

\caption[]{The above table contains the simulation name combined with the maximum
  refinement level$^{a}$, the redshift$^{b}$ at the end of the simulation, the total mass$^{c}$ 
  (gas \& dark matter) at the virial radius ($M_{\odot}$), the virial radius$^{d}$ (kpc), 
  the virial velocity$^{e}$ (km $\mathrm{sec^{-1}}$), the virial temperature$^{f}$ (K), 
  the DM mass$^{g}$ within the virial radius ($M_{\odot}$), the maximum number density$^{h}$ in 
  the halo ($\rm{cm^{-3}}$), the dark matter particle mass$^{i}$ ($M_{\odot}$) and the spatial 
  resolution$^j$ (pc) of the simulation.}

\label{ScalingTable}
\end{table*}


\section{Resolution Dependence of the Outcome of Adaptive Mesh 
Collapse Simulations} 
\label{Scaling}

\subsection{Varying the Resolution}

The maximum spatial (physical) resolution of our fiducial simulations discussed so far 
is $\approx 10^{-3}$ pc and the minimum dark matter particle mass is 
$\rm{M_{DM}} = 8.301 \times 10^2$ \msolar. We investigate now the effect that increasing and 
decreasing the maximum refinement level has on the outcome of our simulations. For this we 
rerun simulations A \& C with different maximum refinement levels.  We have also decreased 
the minimum dark matter particle mass for the highest refinement runs. The parameters of the 
simulations used for this resolution study are shown in Table \ref{ScalingTable}. In Table 
\ref{ScalingTable} simulation A12 refers to simulation A run with a maximum refinement 
level of 12, A18 is simulation A run with a maximum refinement level of 18 (which was the 
fiducial value used in this study), etc. The same notation applies to simulation C. 
Note that the simulations run at a refinement level of 12 are able to run for considerably 
longer as they do not resolve the collapsing structures as early as the higher resolution 
simulations. All simulations were run until the maximum refinement level was reached and 
the simulations were no longer able to properly compute the hydrodynamics due to the lack 
of resolution.
\begin{figure*}
  \centering 
  \begin{minipage}{175mm}      \begin{center}
    \centerline{
      \includegraphics[width=9cm]{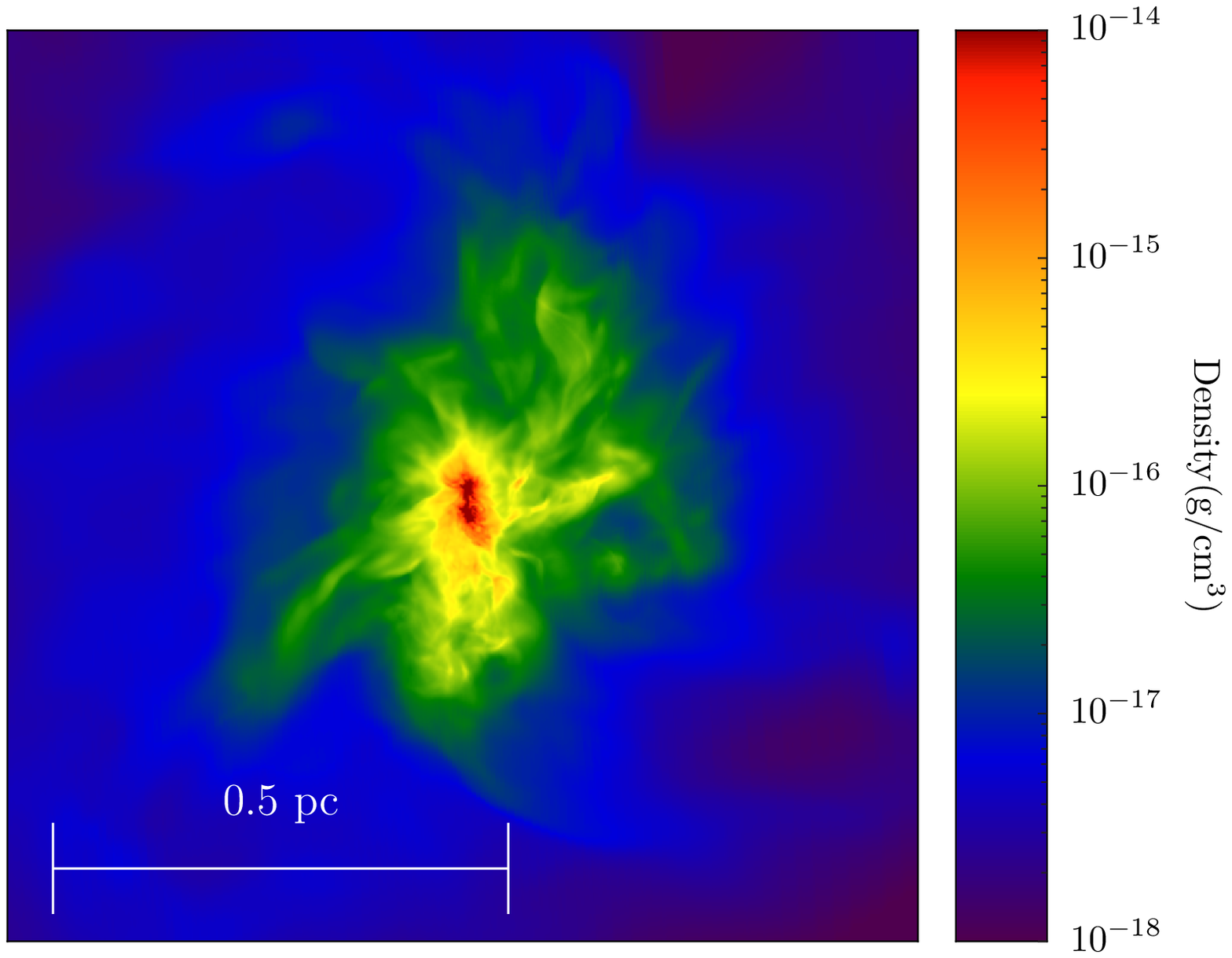}
      \includegraphics[width=9cm]{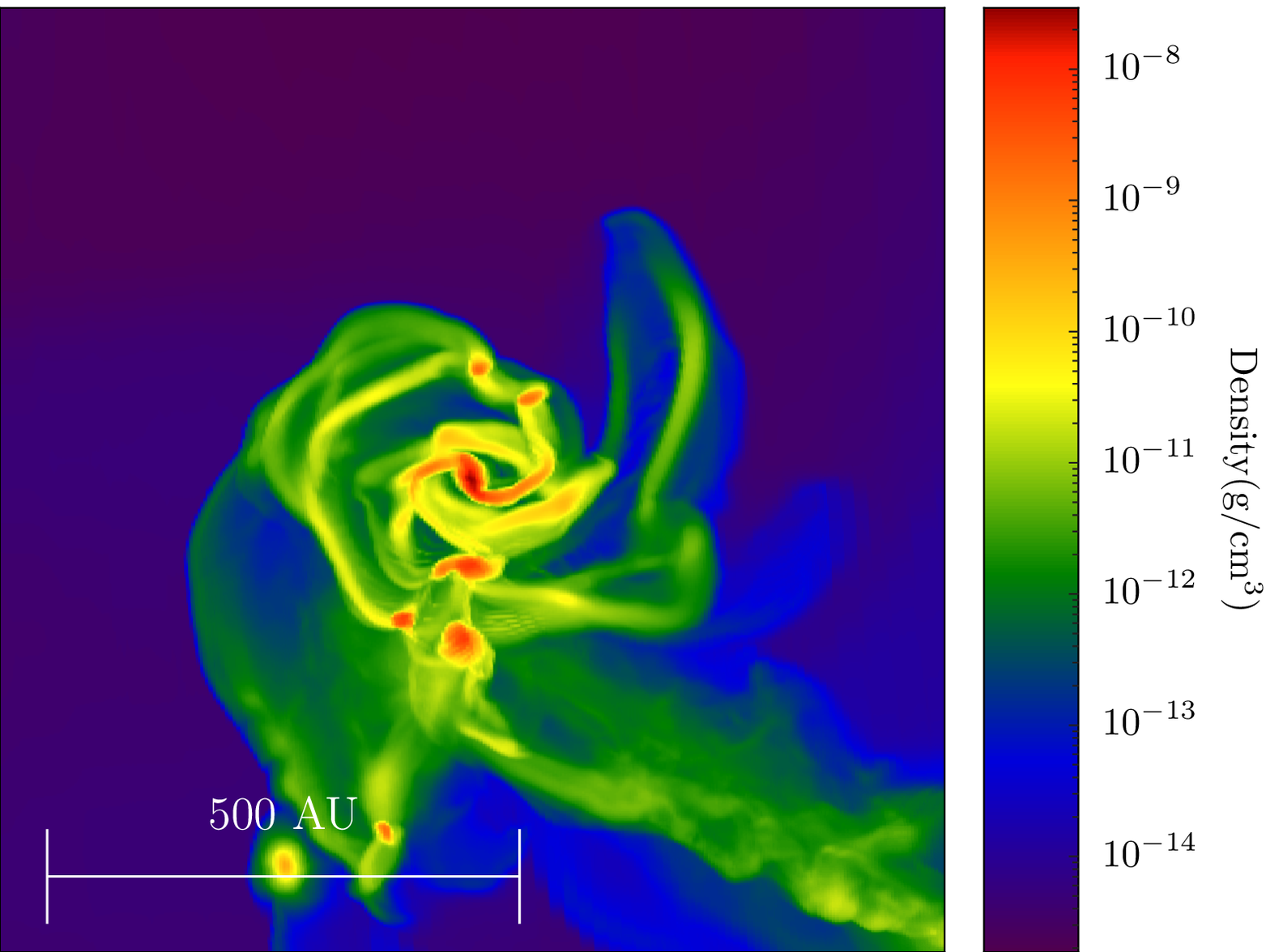}
    }
    \caption[]
    {\label{Halo8_ScaledMultiPlot}
      The above figure shows a density projection centred on the 
      highest density point in Halo C26. The ``up'' vector is chosen to be the
      angular momentum vector and so we are looking down onto the central 
      object. In the left panel the width of the visualisation is 1 pc while on
      the right we have zoomed in onto the densest clump - the width of the panel 
      is 1000 AU. 
}   
   \end{center} \end{minipage}
\end{figure*}


\begin{figure*}
  \centering 
  \begin{minipage}{175mm}      \begin{center}
    \centerline{\includegraphics[width=9cm]{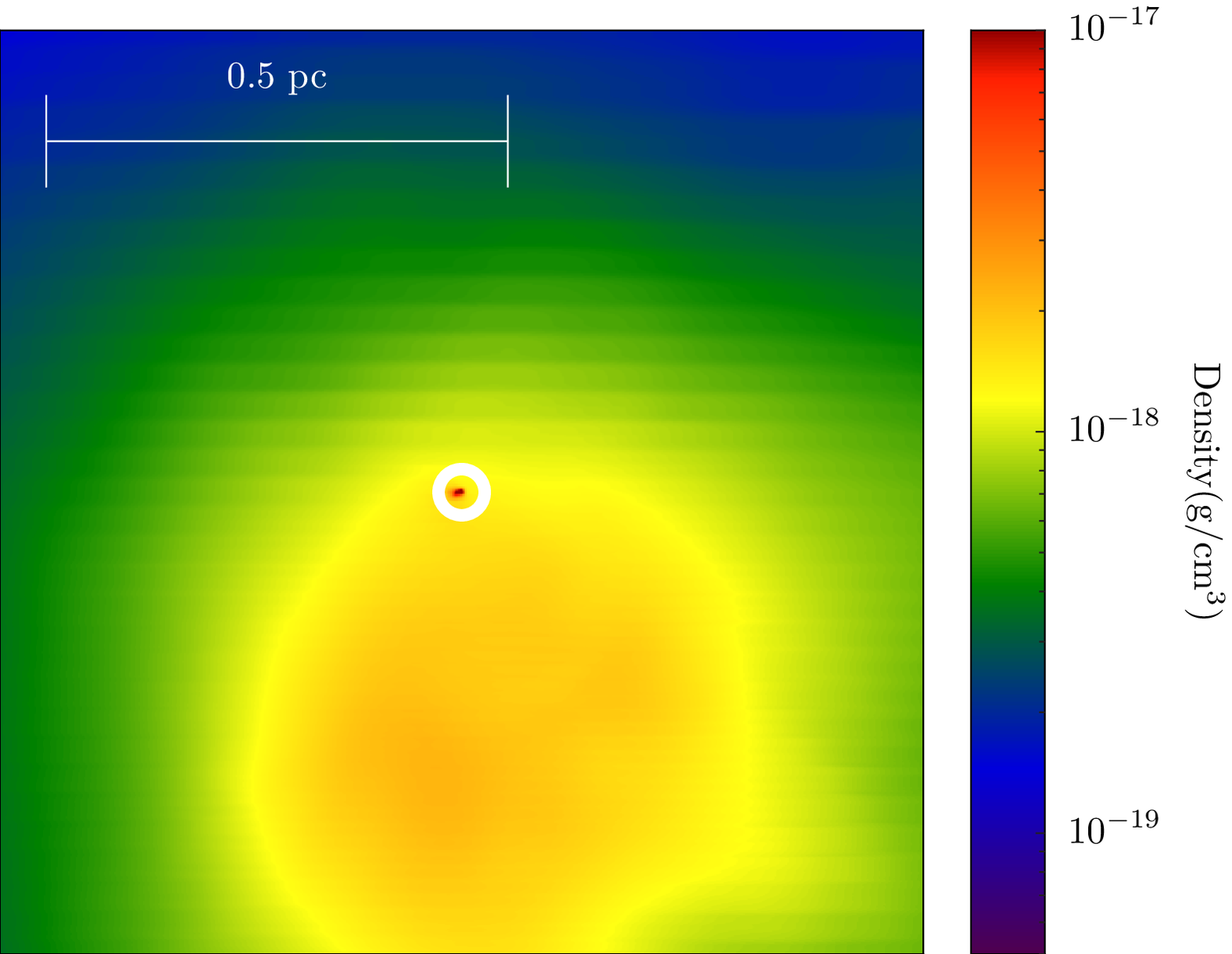}
    \includegraphics[width=9cm]{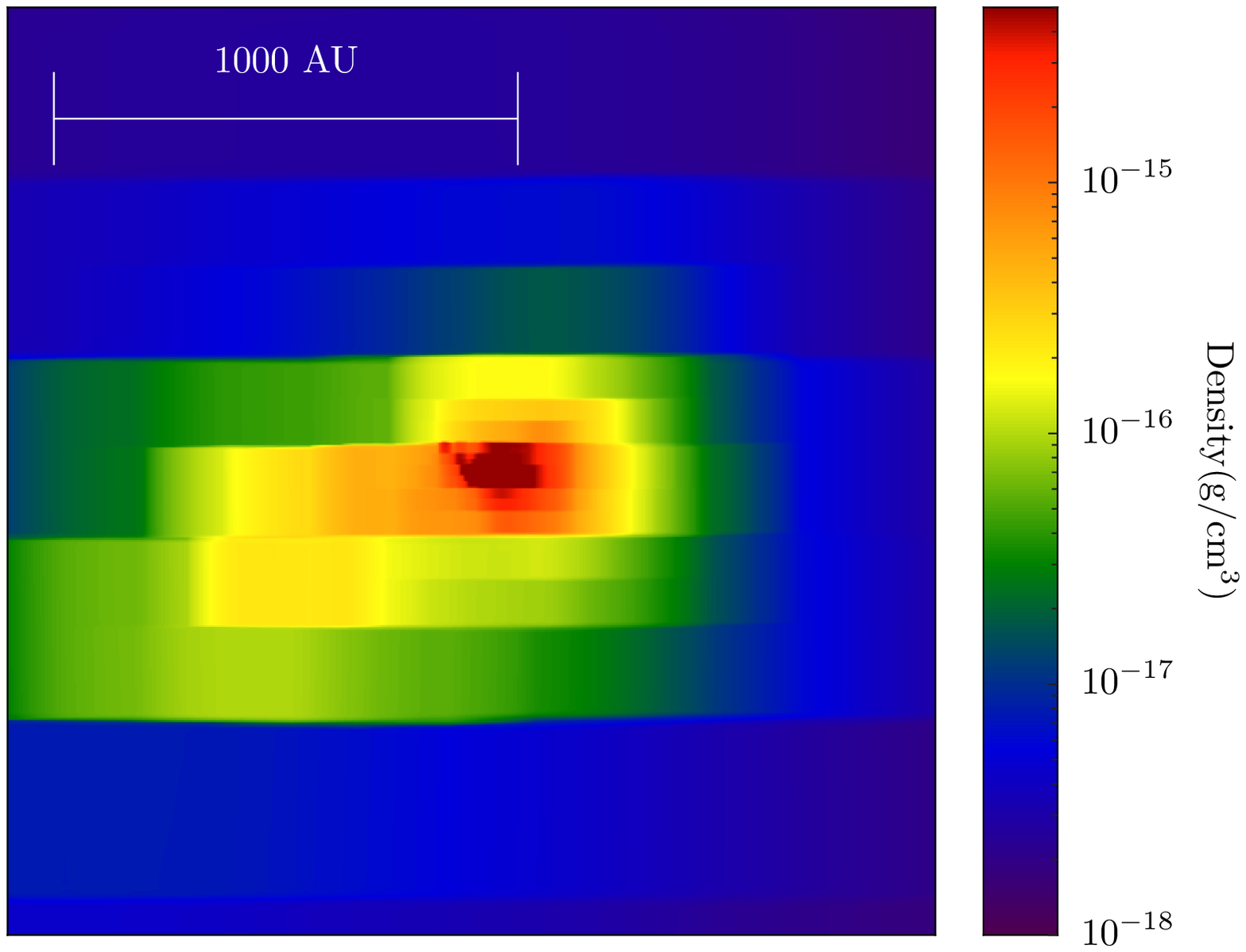}}
    \caption[]
    {\label{Halo2_ScaledMultiPlot}
      The above figure shows a density projection centred on the 
      highest density point in Halo A26. The ``up'' vector is chosen to be the
      angular momentum vector and so we are looking down onto the central 
      object. In the left panel the width of the panel is 1 pc while on
      the right we have zoomed in onto the densest clump - the width of the panel is 
      2000 AU. The zoomed in region is denoted by a white circle in the left hand
      panel.
}   
   \end{center} \end{minipage}
\end{figure*}


\begin{figure*}
  \centering 
  \begin{minipage}{175mm}      \begin{center}
    \centerline{ \psfig{file=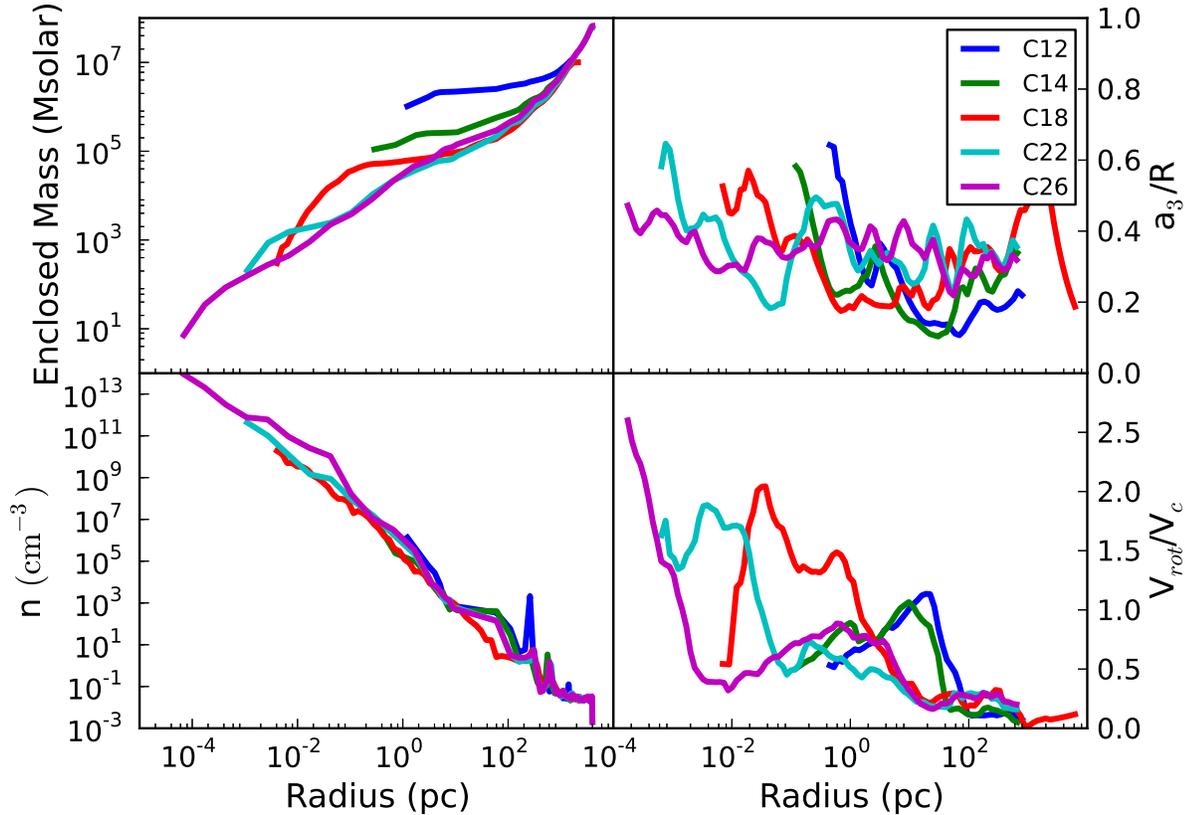,width=18cm}}
    \caption[]
    {\label{Halo8MultiPlot}
       The radial profiles for each simulation in the resolution study of 
       Simulation C.
       {\it Bottom Left Panel:} 
       The density profile against radius. 
       {\it Top Left Panel:}
       The enclosed mass against radius. 
       {\it Top Right Panel:}
       The ratio of the square root of the minimum eigenvalue and radius
       against radius. 
       {\it Bottom Right Panel:}
       The ratio of the rotational velocity ($\rm{V_{rot}}$) and the 
       circular velocity ($\rm{V_{c}}$) plotted against radius.
      
    }   
   \end{center} \end{minipage}
\end{figure*}

\begin{figure*}
  \centering 
  \begin{minipage}{175mm}      \begin{center}
    \centerline{ \psfig{file=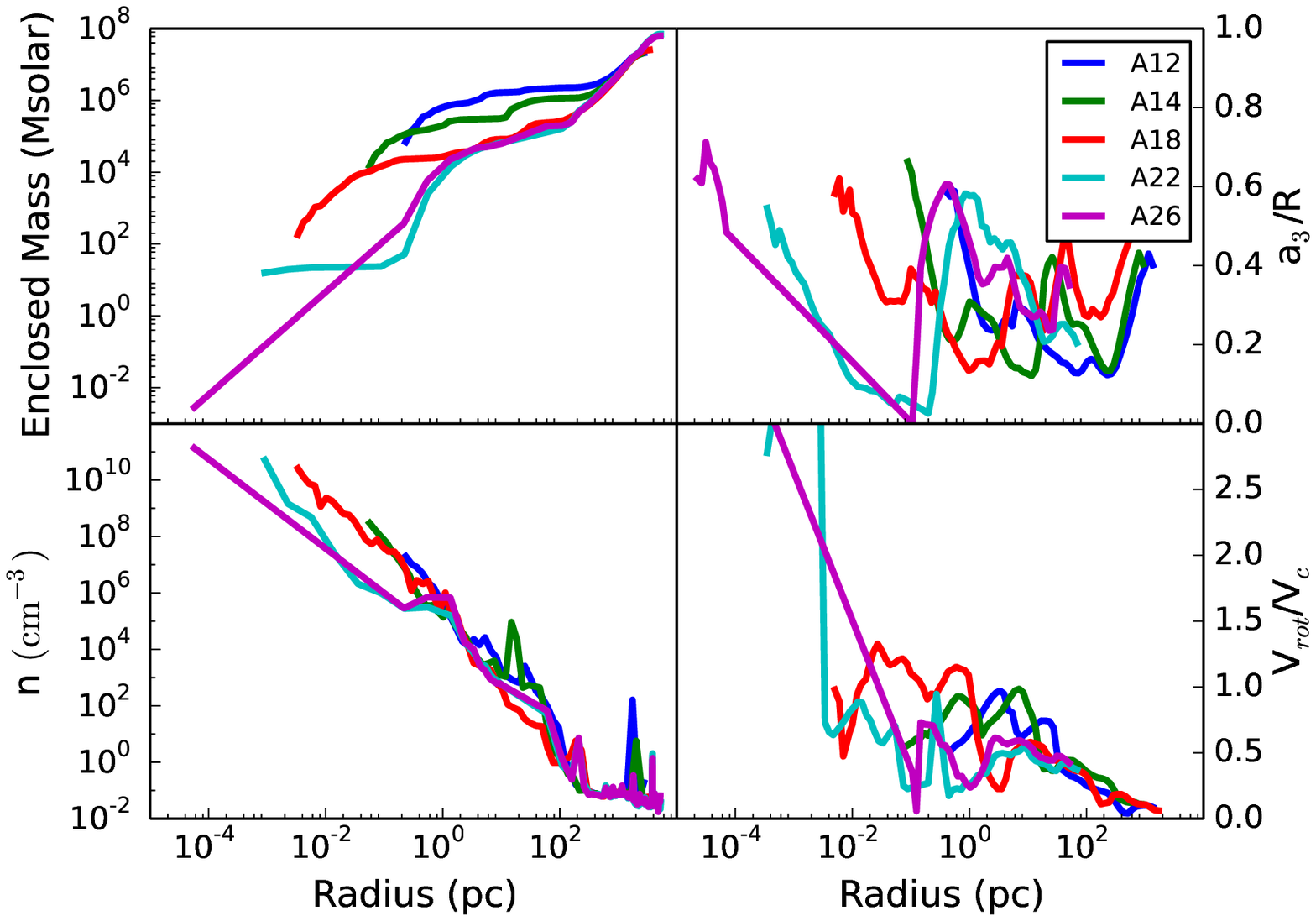,width=18cm}}
    \caption[]
    {\label{Halo2MultiPlot}
     The same as Figure \ref{Halo8MultiPlot} for Simulation A. 
    }   
   \end{center} \end{minipage}
\end{figure*}


\subsection{Splitting of Dark Matter Particles}

When we run the higher resolution simulations in Table \ref{ScalingTable}   
with higher refinement levels we have also decreased the minimum mass 
of the dark matter particles in these simulations. This was to ensure that the 
dark matter particles do not introduce any artificial fragmentation 
in the higher resolution simulations. In the higher resolution runs, those with 
a maximum refinement of $>$ 18, the gas mass resolution becomes substantially higher. 
We initially ran simulations without altering the dark matter particle mass, but found 
that the dark matter particles were aiding fragmentation at small scales in some cases. 
We therefore utilised a modified form of the dark matter splitting algorithm in \enzo - 
which follows the prescription given in \cite{Kitsionas_2002}. The initial 
dark matter particle mass was $\rm{M_{DM}} = 8.301 \times 10^2$ $M_{\odot}$. The 
splitting algorithm splits the dark matter particle into 13 new particles each of 
mass  $\rm{M_{DM}} = 6.385 \times 10^1$ $M_{\odot}$. We did this for simulations 
A22, A26, C22 and C26. Simulations A22 and A26 were split and restarted at 
z = 22 while simulations C22 and C26 were split and restarted at z = 18. In order 
to fully utilise this feature of \enzo we had to modify the \enzo code that 
contains the dark matter splitting algorithm to suit our specific needs. The 
patch was then upstreamed to the \enzo mainline. 

\subsection{Results of the Resolution Study}
\label{Results}
In Figures \ref{Halo8_ScaledMultiPlot} and \ref{Halo2_ScaledMultiPlot} we show 
a density projection of the output at the end of each of the highest resolution  
runs for simulations C and A  (Table \ref{ScalingTable}). The projection is centred on the point 
of highest density in each case. The projection is created using the YT analysis suite 
\citep{YT}. To create the projections the radiative transfer equation along the line of sight 
is integrated by converting field values to emission and absorption values producing a final image.
Fragments are easily identified in this projection.  As the maximum refinement level and therefore 
the maximum resolution of the simulation increases, more and more of the computational resources 
are directed at the densest clump(s) collapsing first. The outer collapsing mass shells 
"freeze out" more quickly with increasing maximum refinement level. This strongly affects the 
resulting morphology. \\
\indent Figure \ref{Halo8_ScaledMultiPlot} shows density projections of the results from our high 
resolution run of simulation C with a maximum refinement level of 26 at two different zoom levels. 
The left panel has a size of 1 pc while the right panel shows the same time output, but zoomed in by 
a factor of $\approx 200$. The right hand panel displays clear evidence for the onset of 
fragmentation within a marginally stable disc with a mass of a few times $10^2$ \msolarc, which 
we will explore in more detail in subsection \ref{Clumps}.  Similarly, Figure 
\ref{Halo2_ScaledMultiPlot} shows the density projection for simulation A26. Here a single dense 
clump of gas (marked by the white circle) fragments at the edge of the main structure 
(Mass $\sim 10^4$ \msolar) and collapses to high density early before the rest of the gas can 
evolve further. \enzo follows this small fragment at the expense of the evolution of the outer 
mass shells. The width of the left panel is 1 pc. The zoomed in visualisation on the right, with 
a width of 0.01 pc, is centred on this clump which has a mass of $\approx 10$ \msolar 
and is an order of magnitude less massive than the clump which collapses at the 
centre of simulation C26. \\
\indent In Figure \ref{Halo8MultiPlot} we show how the radial profiles of enclosed mass, density,
disc thickness and ratio of rotational to circular velocity are affected
by the choice of maximum refinement level.  As expected with increased refinement level the 
collapse can be followed to higher density. The maximum density reached in simulation C26 is 
$> 1 \times 10^{16} \rm{cm^{-3}}$, similar to that in the highest resolution runs presented in 
\cite{Latif_2013c}. Note that for the highest densities the gas should have become optically 
thick to Thompson scattering and our assumption that the gas is optically thin to cooling 
radiation breaks down even for continuum radiation (see section \S \ref{limits} for more 
discussion). The top left panel of Figure \ref{Halo8MultiPlot} shows the enclosed mass against 
radius. C12, C14, C18 and to a somewhat smaller extent C22 all show plateaus in the radial enclosed 
mass profile which indicate that the collapse becomes marginally rotationally supported.  
The enclosed mass where this happens decreases, however, with increasing resolution as the code
follows an increasingly smaller fraction of the gas at increasing resolution. Note also 
the absence of such a plateau in simulation C26. In C26 the time steps have become so 
short that the settling into rotational support cannot be tracked for any of the outer mass 
shells. In C26 a disk does not form at all, because it does not have adequate time to do so.  
The dips in the top right panel of Figure \ref{Halo8MultiPlot} for C12, C14, C18 and C22  
showing the ratio of the eigenvalue $a_3$ of the inertia tensor to the radius as a proxy for 
disk thickness suggests that indeed a (fat) disk has formed in the lower resolution simulations.  
C26 shows no obvious dip. 
In the bottom right panel of Figure \ref{Halo8MultiPlot} C12 and C14 show that marginal 
rotational support is achieved at scales of $\sim 10 - 20$ pc, C18 and C22 show strong 
rotational support at scales of $\sim$ a few times $10^{-1}$ and $\sim$ a few times $10^{-2}$ pc 
respectively. C26 shows no sign of rotational support - the ratio goes only above 1.0 near the 
resolution limit and is therefore not indicative of a settled disk. \\
\indent Figure \ref{Halo2MultiPlot} shows the radial profiles for simulation A. Again with 
increasing resolution higher densities are reached, but as we can clearly see here it is not 
necessarily the central regions of the halo which are collapsing first. The decoupling of the 
small off-centre fragment from the rest of the gas has a noticeable effect on the profile in 
simulations A22 and A26. However, in A22 the fragment rejoins the collapsing outer structure 
as the simulation progresses. In the top left panel of Figure \ref{Halo2MultiPlot} we show 
the enclosed mass as a function of radius.  A26 shows no sign of disk formation and 
the simulation follows only the collapse of the off-centre clump to ever higher densities at
late times. It is also worth noting that in the right hand panel of Figure \ref{Toomre} 
where we plot the Toomre stability parameter, the disk in simulation A18 is observed to be highly 
gravitationally unstable while the disk in simulation C18 appears to be (marginally) 
gravitationally stable within a scale length or more. This suggests that at a resolution of 
$\sim 1 \times 10^{-3}$ pc (maximum refinement = 18) the disk in simulation A is already highly 
gravitationally unstable, but the resolution of the simulations is not sufficient to 
detect the onset of fragmentation.

\begin{figure*}
  \centering
  \begin{minipage}{175mm}      \begin{center}
      \centerline{ \psfig{file=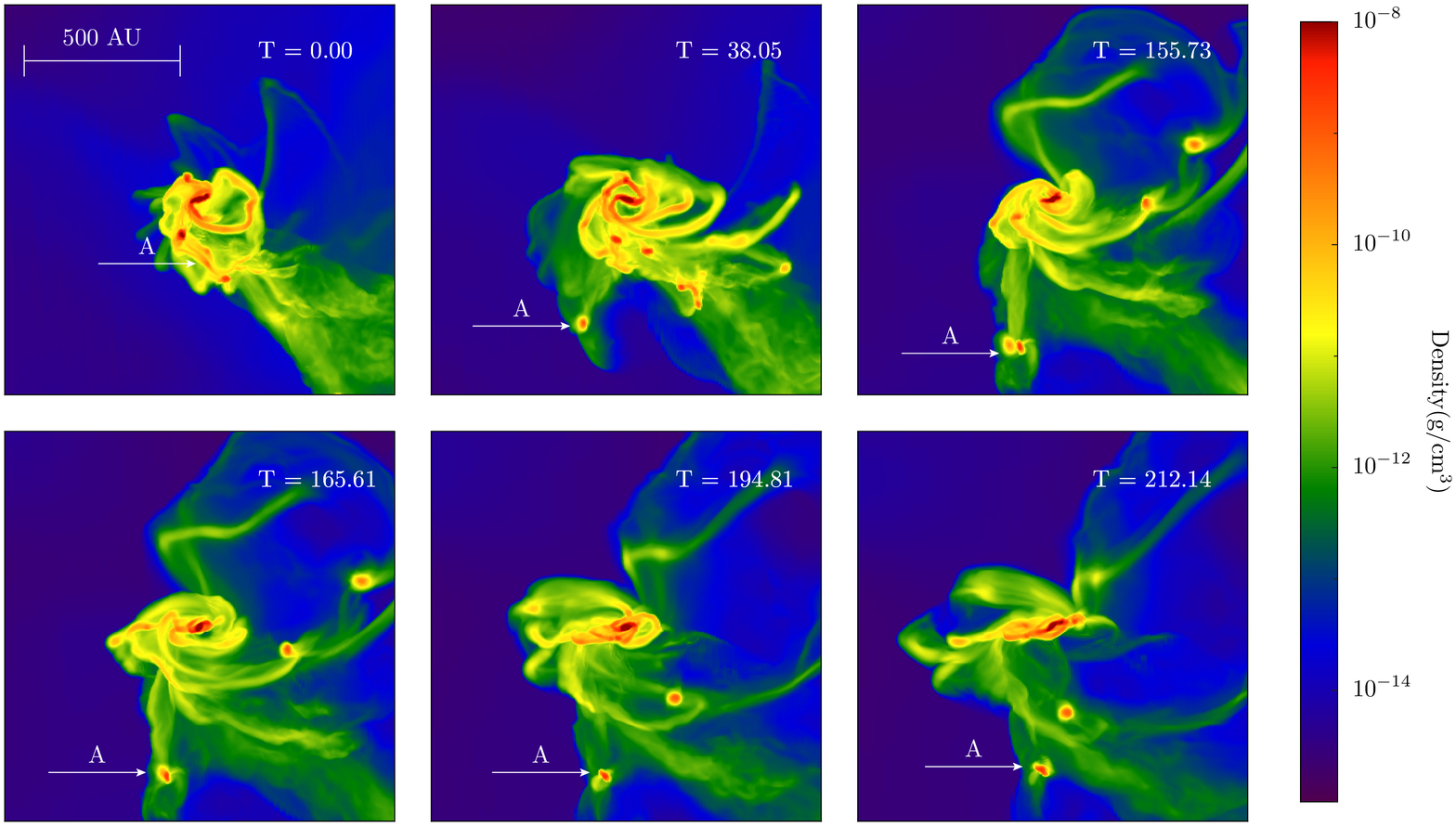,width=18cm}}
      \caption[]
      {\label{ClumpAMultiPlot}
        Time sequence of clump formation. All times are in years. We follow the 
        evolution of a single clump, identified by the letter 'A', in this sequence. 
        The clump is first identified visually in the top left panel 
        and we designate the time here as T = 0 yrs. The clump is then 
        followed over several dynamical times. At approximately T = 155 years the clump 
        merges with another clump. This event causes the clump mass to exceed the 
        Jeans mass. See text for further details. 
      }   
  \end{center} 
  \end{minipage}
\end{figure*}


\begin{figure*}
  \centering 
  \begin{minipage}{175mm}      \begin{center}
    \centerline{ \psfig{file=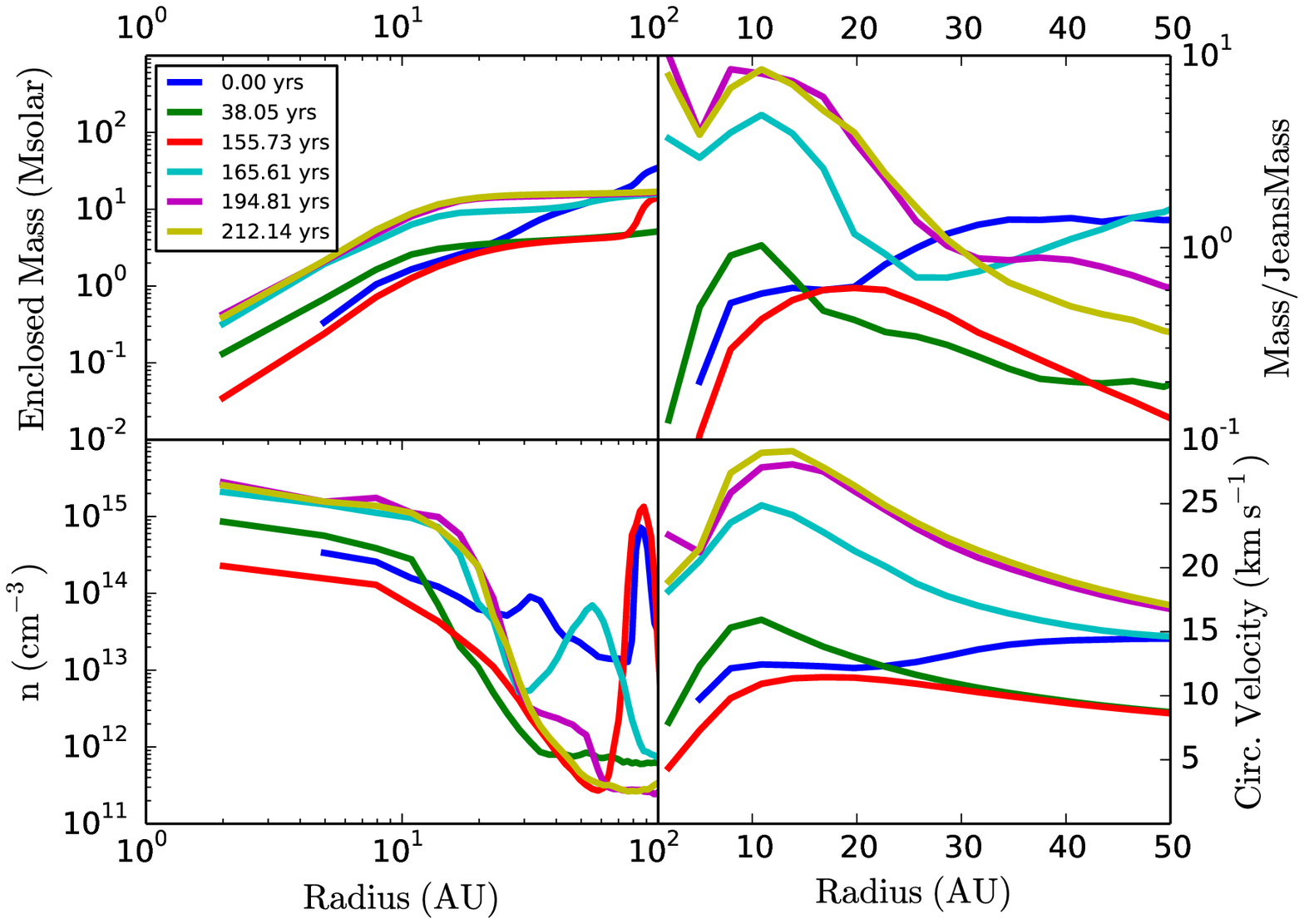,width=18cm}}
    \caption[]
    {\label{ClumpACharPlot}
      The radial profiles for clump A, as identified in Figure \ref{ClumpAMultiPlot}.
      {\it Bottom Left Panel:} 
      The density profile against radius. 
      {\it Top Left Panel:}
      The enclosed mass against radius. 
      {\it Top Right Panel:}
      The ratio of the enclosed mass and the Jeans mass of the clump
      against radius. 
      {\it Bottom Right Panel:}
      The circular velocity plotted against radius.
    }   
   \end{center} \end{minipage}
\end{figure*}

\subsection{Choosing the Right Resolution}
The above results demonstrate that choosing the correct maximum resolution for 
collapse simulations is of crucial importance. Both choosing a resolution that is too low and 
equally choosing a resolution that is too high can lead to a misleading interpretation of results. 
Our lower resolution simulations, C14 for example, provide tentative evidence that rotational 
support may extend out to $\sim 10$ pc or more, but they lack the required resolution to probe the
dynamics at sub-parsec scales. In C26 we showed that at very high resolution a rotationally 
supported disk does not form. This is because the simulation was unable to follow the outer 
mass shells for the required dynamical time in order for the disks to form. The dynamical 
time required to form the rotationally supported disk in C18 is greater than 10 Myrs, however, 
C26 only evolves gas above a similar density for the order of 1 to 2 Myrs. The high resolution 
runs did however, indicate that fragmentation is common at sub-parsec scales in agreement with 
the results found by \cite{Latif_2013c}. In addition, running simulations that 
closely resembles our highest resolution simulations \cite{Latif_2013c} also found that 
a disk does not form. In both cases, the reason is most likely that the simulation does not run 
for the required dynamical time of the outer shells rather than the fact that a disk does not form 
at all. This is an inherent limitation of high resolution AMR simulations that can only be 
overcome by running simulations at varying resolutions. \\ \indent
Increasing the resolution in simulations  without a feedback mechanism which will set a 
characteristic resolution scale, means that an AMR code will follow the densest fluctuation at the 
expense of lower density gas. Feedback mechanisms such as turbulence \citep{Latif_2013c}, 
magnetic fields \citep{Latif_2013e}, or ionising radiation from a previous source 
naturally provide a mechanism to prevent or reduce fragmentation. In the case of such a 
feedback mechanism high resolution simulations are extremely useful, because the simulation is 
able to track the dynamics of the in-falling gas at both very small and intermediate scales. The 
feedback sets a characteristic resolution scale by acting against density fluctuations at very 
small scales. Without such feedback mechanisms converged numerical results are impossible to reach 
in this case. This is due to the nature of the AMR simulations conducted. Multiple simulations at 
varying resolutions are therefore required to ascertain the complete picture.

\subsection{Limitations of the Six Species Cooling Model}
\label{limits}
As already discussed our simulations have included only the six species chemical model (
${\rm H}, {\rm H}^+, {\rm He}, {\rm He}^+,  {\rm He}^{++}, {\rm e}^-$). We have neglected the 
presence of \molH in our simulations and have thus implicitly assumed a dissociating Lyman-Werner 
background capable of destroying all the \molH within our halo. As discussed in \S \ref{H2} 
a background capable of keeping a halo \molH free is certainly plausible at this redshift for 
selected haloes in the Universe, at least in the initial stages of the collapse. However, at the 
densities reached in our high resolution simulations it is likely that \molH will be formed, 
through both the usual $\rm{H^-}$ and $\rm{H^+}$ channels and via three body reactions 
involving $\rm{H}$ \citep{Palla_1983}, at the very centre of our collapsing object, possibly at a 
rate faster than the dissociation timescale. \cite{Shang_2010} ran several simulations with 
varying \molH dissociating backgrounds. Their strongest fields, $\gtrsim 10^3 J_{21}$, effectively 
reduce the \molH fraction to $\lesssim 10^{-8}$, which is several orders of magnitude below the 
level required to influence the gas thermodynamics. However, the densities reached in their 
simulations were significantly below those reached in our high resolution runs. 
Radiative transfer methods will be required to accurately model this environment as we exit the 
regime in which the optically thin approximation holds. \\
\indent In simulation C26, for example, the highest densities reached are $\rm{n} \sim 10^{16}\ 
\rm{cm^{-3}}$. As already noted in \S \ref{Results} we have now moved outside the realm in which the 
optically thin approximation is valid and hence we are no longer accurately modelling the cooling 
processes which are activated at these high densities (e.g. line trapping of cooling radiation, 
Thomson scattering). This will have the knock-on effect of impacting the subsequent cooling and 
possibly the evolution of the central object. In a future study we will investigate in detail the
effects of a Lyman-Werner dissociating background, fully taking into account the effect
a dissociating flux has on the \molH cooling and therefore on the gas thermodynamics.

\subsection{The Onset of Fragmentation and the Formation of Bound Clumps} 
\label{Clumps}

In the absence of metal and molecular hydrogen cooling fragmentation is strongly suppressed 
in collapsing dark matter haloes at the atomic cooling threshold \citep{Oh_2002, Bromm_2003,
Lodato_2006, Spaans_2006, Begelman_2006, Schleicher_2010}. This is because the cooling 
effectiveness of atomic hydrogen lines drops below $ \sim 10000\ K$. The temperature remains 
at close to 7000K for gas outside of the core of the halo, but can drop to between 3000 K and 
5000 K at the very centre of the highest density regions. As the collapse becomes 
marginally rotationally supported and the gas settles into a disc like structure, our simulations 
begin, however, to show the onset of significant fragmentation and the formation of clumps.  
In Figure \ref{ClumpAMultiPlot} we show a time series of outputs at fixed density and fixed size 
of our simulation C26. Several clumps are clearly noticeable.  Note that the maximum circular 
velocity of the marginally stable disc at this stage is $\gtrsim 40$ km/s. The temperature of 
the gas is therefore well below the ``virial temperature'' of the \emph{disk} which 
may explain the onset of fragmentation at this stage. Fragmentation may also be aided by the 
interaction of the gas that is flung out with the gas flowing inwards from larger radii 
(see \citealt{Bonnell_1994b} for a similar discussion in a different context).\\
\indent The clumps have typical masses of 5 - 20 \msolar and could therefore be potential formation 
sites of stars.  Let us now have a closer look at clump A - identified in 
Figure \ref{ClumpAMultiPlot}. The time at which the clump is first visually identified is 
selected as T = 0 and shown in the top left panel. The clump merges with another clump in the 
third picture of the sequence and the two clumps coalesce before circling back towards the dynamical 
centre. The physical properties of clump A are shown in Figure \ref{ClumpACharPlot}. Using the clump 
finding algorithm \citep{Smith_2009} built into YT we track the clump over several 
dynamical times. In the bottom left panel of Figure \ref{ClumpACharPlot} we plot the density 
against radius for the clump. As the clump moves away from the central density and 
collapses the profile steepens significantly between T = 0 yrs and T = 212 yrs. 
The radius of the clump is approximately 20 AU. In the top left panel we show the 
enclosed mass profile. The enclosed mass  grows  approximately linearly with the radius out
to about 20 AU.  The bottom right hand panel shows  the circular velocity of the clump against 
radius. The circular velocity increases strongly between 20 and 40 AU, reaching a maximum value 
of $\approx 25 - 30 \ \rm{km\ s^{-1}}$. The clump is thus strongly bound. In the top right hand 
panel we plot the ratio of the enclosed mass divided by the Jeans mass. Initially the clump is 
Jeans stable, but as the clump evolves, and through its merger with another clump, 
the clump grows in mass, becomes Jeans unstable, and collapses.

\subsection{Predicting the Further Evolution}

It is very difficult to follow the long-term evolution of the bulk of the gas in the 
haloes at high resolution due to the prohibitively short dynamical time scales. As we have
discussed earlier, there are, however, clear signs for the onset of fragmentation in the 
simulations, despite the absence of metal and molecular hydrogen cooling. The long term fate of 
these fragments is uncertain at this point, some of them appear to dissolve quickly, but some 
of them appear to become strongly bound and are likely to persist and form stars. Predicting 
the final outcome of these simulations even for the simplified case of no radiative and 
supernova feedback and no optical depth effects is thus not yet possible and will probably have 
to involve simulations making use of sink particles. Judging from our current simulations, 
it appears that it is rather unlikely that most of the gas at the centre of these haloes will 
rapidly accrete onto a single star/black hole and that a dense star cluster may form instead which  
then evolves further \citep{Regan_2009b}.  However, the inclusion of radiative transfer and 
feedback effects may well again change the outcome not only quantitatively, but 
also qualitatively. 
 
\section{Conclusions}
\label{conclusions}

We have performed a suite of \enzo AMR simulations of the collapse of gas in DM haloes with 
virial temperatures just above the atomic cooling threshold. As in previous simulations by us 
and other authors the highly turbulent gas attains an isothermal density profile during the 
collapse. The gas looses angular momentum efficiently due to angular momentum separation and 
cancellation and settles towards the centre at a significant fraction of the free-fall velocity. 
Furthermore, within our suite of simulations, we ran an extended set of simulations at 
varying maximum refinement level. We found that the gas becomes marginally rotationally supported at 
characteristic radii, settles and forms a thick often gravitationally unstable disc. 
The mass at which the onset of rotational support can be studied by such AMR simulations 
is set by the refinement level and decreases with increasing maximum refinement level. At 
the highest refinement levels for which we ran our simulations the dynamical times become 
too short to study the onset of rotational support at any radius. The interpretation of these 
AMR simulations requires therefore great care. Similar to what was found in the simulations 
by \cite{Latif_2013a, Latif_2013c, Latif_2013d} we found that in our highest resolution 
simulations (maximum refinement level 26) the gas reaches number densities as high as 
$10^{16} \rm{cm^{-3}}$ and is prone to fragmentation. The fragments have masses of a few tens 
of solar masses and radii of 10-20 AU. These dense often strongly bound clumps with 
circular velocities $>10$ km/s forming in the highest resolution simulations are, however, 
not necessarily located at the very centre of the halo. \\ \indent
They are probably best interpreted as the onset of fragmentation within the marginally stable,
highly self-gravitating, discs which appear to be a generic outcome of these simulations. 
Predicting the further evolution of the collapse is not yet possible with the current setup, 
but the formation of a dense stellar cluster, perhaps as an intermediate stage to the formation 
of a massive seed black hole is a possible outcome. Future progress will require the inclusion of 
feedback effects due to the radiation of massive stars and supernovae and probably also the 
employment of appropriately chosen sink particles in order to follow the evolution of the 
unstable discs over longer timescales. 

\section*{Acknowledgements}
J.A.R. and P.H.J. acknowledge the support of the Magnus Ehrnrooth Foundation and the Research
Funds of the University of Helsinki. J.A.R. and M.G.H. further acknowledge support from the 
FP7 ERC Advanced Grant Emergence-320596. The numerical simulations were performed on facilities 
hosted by the CSC -IT Center for Science in Espoo, Finland, which are financed by the 
Finnish ministry of education and the UK National Cosmology Supercomputer - \cosmos - which is 
part of DiRAC - the UK distributed HPC Facility. The freely available astrophysical analysis 
code YT \citep{YT} was used to construct numerous plots within this paper. The authors would 
like to express their gratitude to Matt Turk et al. for an excellent software package. 
J.A.R. would also like to thank Jorma Harju for useful discussions on molecular hydrogen 
dissociation. M.G.H thanks Jim Pringle for helpful discussions. We are grateful to the referee 
for a detailed report which improved the paper.

\bibliographystyle{mn2e_w}
\bibliography{/home/regan/Research/Papers/WorkingCopy/AstroPH/Updated/mybib}

\begin{thebibliography}{105}
\providecommand{\natexlab}[1]{#1}

\bibitem[{{Agarwal} et~al.(2012){Agarwal}, {Khochfar}, {Johnson}, {Neistein},
  {Dalla Vecchia} \& {Livio}}]{Agarwal_2012}
{Agarwal} B., {Khochfar} S., {Johnson} J.~L., {Neistein} E., {Dalla Vecchia}
  C., {Livio} M., 2012, \mnras, 425, 2854

\bibitem[{{Ahn} et~al.(2009){Ahn}, {Shapiro}, {Iliev}, {Mellema} \&
  {Pen}}]{Ahn_2009}
{Ahn} K., {Shapiro} P.~R., {Iliev} I.~T., {Mellema} G., {Pen} U.~L., 2009,
  \apj, 695, 1430

\bibitem[{{Aykutalp} et~al.(2013){Aykutalp}, {Wise}, {Meijerink} \&
  {Spaans}}]{Aykutalp_2013}
{Aykutalp} A., {Wise} J.~H., {Meijerink} R., {Spaans} M., 2013, \apj, 771, 50

\bibitem[{{Begelman}(2001)}]{Begelman_2001}
{Begelman} M.~C., 2001, \apj, 551, 897

\bibitem[{{Begelman}(2008)}]{Begelman_2008b}
{Begelman} M.~C., 2008, in B.W. {O'Shea}, A.~{Heger}, eds, First Stars III.
  American Institute of Physics Conference Series, Vol. 990, pp. 489--493

\bibitem[{{Begelman} \& {Rees}(1978)}]{Begelman_1978}
{Begelman} M.~C., {Rees} M.~J., 1978, \mnras, 185, 847

\bibitem[{{Begelman} et~al.(1984){Begelman}, {Blandford} \&
  {Rees}}]{Begelman_1984}
{Begelman} M.~C., {Blandford} R.~D., {Rees} M.~J., 1984, Reviews of Modern
  Physics, 56, 255

\bibitem[{{Begelman} et~al.(2006){Begelman}, {Volonteri} \&
  {Rees}}]{Begelman_2006}
{Begelman} M.~C., {Volonteri} M., {Rees} M.~J., 2006, \mnras, 370, 289

\bibitem[{{Begelman} et~al.(2008){Begelman}, {Rossi} \&
  {Armitage}}]{Begelman_2008}
{Begelman} M.~C., {Rossi} E.~M., {Armitage} P.~J., 2008, \mnras, 387, 1649

\bibitem[{{Bellovary} et~al.(2013){Bellovary}, {Brooks}, {Volonteri},
  {Governato}, {Quinn} \& {Wadsley}}]{Bellovary_2013}
{Bellovary} J., {Brooks} A., {Volonteri} M., {Governato} F., {Quinn} T.,
  {Wadsley} J., 2013, \apj, 779, 136

\bibitem[{{Berger} \& {Colella}(1989)}]{Berger_1989}
{Berger} M.~J., {Colella} P., 1989, Journal of Computational Physics, 82, 64

\bibitem[{{Berger} \& {Oliger}(1984)}]{Berger_1984}
{Berger} M.~J., {Oliger} J., 1984, Journal of Computational Physics, 53, 484

\bibitem[{{Binney} \& {Tremaine}(2008)}]{Binney_2008}
{Binney} J., {Tremaine} S., 2008, {Galactic Dynamics: Second Edition}.
  Princeton University Press

\bibitem[{{Bonnell}(1994)}]{Bonnell_1994b}
{Bonnell} I.~A., 1994, \mnras, 269, 837

\bibitem[{{Bromm} \& {Loeb}(2003)}]{Bromm_2003}
{Bromm} V., {Loeb} A., 2003, \apj, 596, 34

\bibitem[{{Bromm} et~al.(2009){Bromm}, {Ferrara} \& {Heger}}]{Bromm_2009}
{Bromm} V., {Ferrara} A., {Heger} A., 2009, {First stars: formation, evolution
  and feedback effects}, Cambridge University Press. p. 180

\bibitem[{{Bryan} \& {Norman}(1995)}]{Bryan_1995b}
{Bryan} G.~L., {Norman} M.~L., 1995, Bulletin of the American Astronomical
  Society, 27, 1421

\bibitem[{{Bryan} \& {Norman}(1997)}]{Bryan_1997}
{Bryan} G.~L., {Norman} M.~L., 1997, in ASP Conf. Ser. 123: Computational
  Astrophysics; 12th Kingston Meeting on Theoretical Astrophysics. pp. 363--+

\bibitem[{{Cen} \& {Riquelme}(2008)}]{Cen_2008}
{Cen} R., {Riquelme} M.~A., 2008, \apj, 674, 644

\bibitem[{{Choi} et~al.(2013){Choi}, {Naab}, {Ostriker}, {Johansson} \&
  {Moster}}]{Choi_2013}
{Choi} E., {Naab} T., {Ostriker} J.~P., {Johansson} P.~H., {Moster} B.~P.,
  2013, ArXiv e-prints 1308.3719

\bibitem[{{Clark} et~al.(2011){Clark}, {Glover}, {Klessen} \&
  {Bromm}}]{Clark_2011}
{Clark} P.~C., {Glover} S.~C.~O., {Klessen} R.~S., {Bromm} V., 2011, \apj, 727,
  110

\bibitem[{{Costa} et~al.(2013){Costa}, {Sijacki}, {Trenti} \&
  {Haehnelt}}]{Costa_2013}
{Costa} T., {Sijacki} D., {Trenti} M., {Haehnelt} M., 2013, ArXiv e-prints
  1307.5854

\bibitem[{{Debuhr} et~al.(2011){Debuhr}, {Quataert} \& {Ma}}]{Debuhr_2011}
{Debuhr} J., {Quataert} E., {Ma} C.~P., 2011, \mnras, 412, 1341

\bibitem[{{Di Matteo} et~al.(2005){Di Matteo}, {Springel} \&
  {Hernquist}}]{DiMatteo_2005}
{Di Matteo} T., {Springel} V., {Hernquist} L., 2005, \nat, 433, 604

\bibitem[{{Dijkstra} et~al.(2008){Dijkstra}, {Haiman}, {Mesinger} \&
  {Wyithe}}]{Dijkstra_2008}
{Dijkstra} M., {Haiman} Z., {Mesinger} A., {Wyithe} J.~S.~B., 2008, \mnras,
  391, 1961

\bibitem[{{Dubois} et~al.(2012){Dubois}, {Pichon}, {Haehnelt}, {Kimm}, {Slyz},
  {Devriendt} \& {Pogosyan}}]{Dubois_2012}
{Dubois} Y., {Pichon} C., {Haehnelt} M., {Kimm} T., {Slyz} A., {Devriendt} J.,
  {Pogosyan} D., 2012, \mnras, 423, 3616

\bibitem[{{Efstathiou} et~al.(1985){Efstathiou}, {Davis}, {White} \&
  {Frenk}}]{Efstathiou_1985}
{Efstathiou} G., {Davis} M., {White} S.~D.~M., {Frenk} C.~S., 1985, \apjs, 57,
  241

\bibitem[{{Fabian}(2012)}]{Fabian_2012}
{Fabian} A.~C., 2012, \araa, 50, 455

\bibitem[{{Fan} et~al.(2006){Fan}, {Carilli} \& {Keating}}]{Fan_2006}
{Fan} X., {Carilli} C.~L., {Keating} B., 2006, \araa, 44, 415

\bibitem[{{Fan}(2004)}]{Fan_2004}
{Fan} X.~e.~a., 2004, \aj, 128, 515

\bibitem[{{Federrath} et~al.(2011){Federrath}, {Sur}, {Schleicher}, {Banerjee}
  \& {Klessen}}]{Federrath_2011}
{Federrath} C., {Sur} S., {Schleicher} D.~R.~G., {Banerjee} R., {Klessen}
  R.~S., 2011, \apj, 731, 62

\bibitem[{{Ferrarese} \& {Merritt}(2000)}]{Ferrarese_2000}
{Ferrarese} L., {Merritt} D., 2000, \apjl, 539, L9

\bibitem[{{Gebhardt} et~al.(2000)}]{Gebhardt_2000}
{Gebhardt} K. et~al., 2000, \apjl, 539, L13

\bibitem[{{Greif} et~al.(2007){Greif}, {Johnson}, {Bromm} \&
  {Klessen}}]{Greif_2007}
{Greif} T.~H., {Johnson} J.~L., {Bromm} V., {Klessen} R.~S., 2007, \apj, 670, 1

\bibitem[{{Greif} et~al.(2008){Greif}, {Johnson}, {Klessen} \&
  {Bromm}}]{Greif_2008}
{Greif} T.~H., {Johnson} J.~L., {Klessen} R.~S., {Bromm} V., 2008, \mnras, 387,
  1021

\bibitem[{{Greif} et~al.(2012)}]{Greif_2012}
{Greif} T.~H., {Bromm} V., {Clark} P.~C., {Glover} S.~C.~O., {Smith} R.~J.,
  {Klessen} R.~S., {Yoshida} N., {Springel} V., 2012, \mnras, 424, 399

\bibitem[{{Haehnelt} \& {Kauffmann}(2000)}]{Haehnelt_2000}
{Haehnelt} M.~G., {Kauffmann} G., 2000, \mnras, 318, L35

\bibitem[{{Haiman}(2006)}]{Haiman_2006}
{Haiman} Z., 2006, New Astronomy Review, 50, 672

\bibitem[{{Haiman}(2013)}]{Haiman_2012}
{Haiman} Z., 2013, in T.~{Wiklind}, B.~{Mobasher}, V.~{Bromm}, eds,
  Astrophysics and Space Science Library. Astrophysics and Space Science
  Library, Vol. 396, p. 293

\bibitem[{{Haiman} \& {Bryan}(2006)}]{Haiman_2006b}
{Haiman} Z., {Bryan} G.~L., 2006, \apj, 650, 7

\bibitem[{{Haiman} \& {Holder}(2003)}]{Haiman_2003}
{Haiman} Z., {Holder} G.~P., 2003, \apj, 595, 1

\bibitem[{{Haiman} \& {Loeb}(2001)}]{Haiman_2001}
{Haiman} Z., {Loeb} A., 2001, \apj, 552, 459

\bibitem[{{Haiman} et~al.(2000){Haiman}, {Abel} \& {Rees}}]{Haiman_2000}
{Haiman} Z., {Abel} T., {Rees} M.~J., 2000, \apj, 534, 11

\bibitem[{{Hockney} \& {Eastwood}(1988)}]{Hockney_1988}
{Hockney} R.~W., {Eastwood} J.~W., 1988, {Computer simulation using particles}.
  Bristol: Hilger, 1988

\bibitem[{{Hopkins}(2013)}]{Hopkins_2012}
{Hopkins} P.~F., 2013, \mnras, 430, 1653

\bibitem[{{Johansson} et~al.(2009{\natexlab{a}}){Johansson}, {Burkert} \&
  {Naab}}]{Johansson_2009b}
{Johansson} P.~H., {Burkert} A., {Naab} T., 2009{\natexlab{a}}, \apjl, 707,
  L184

\bibitem[{{Johansson} et~al.(2009{\natexlab{b}}){Johansson}, {Naab} \&
  {Burkert}}]{Johansson_2009a}
{Johansson} P.~H., {Naab} T., {Burkert} A., 2009{\natexlab{b}}, \apj, 690, 802

\bibitem[{{Johnson} \& {Bromm}(2007)}]{Johnson_2007}
{Johnson} J.~L., {Bromm} V., 2007, \mnras, 374, 1557

\bibitem[{{Johnson} et~al.(2011){Johnson}, {Khochfar}, {Greif} \&
  {Durier}}]{Johnson_2011}
{Johnson} J.~L., {Khochfar} S., {Greif} T.~H., {Durier} F., 2011, \mnras, 410,
  919

\bibitem[{{Johnson} et~al.(2013){Johnson}, {Dalla} \&
  {Khochfar}}]{Johnson_2013}
{Johnson} J.~L., {Dalla} V.~C., {Khochfar} S., 2013, \mnras, 428, 1857

\bibitem[{{Kauffmann} \& {Haehnelt}(2000)}]{Kauffmann_2000}
{Kauffmann} G., {Haehnelt} M., 2000, \mnras, 311, 576

\bibitem[{{Kitsionas} \& {Whitworth}(2002)}]{Kitsionas_2002}
{Kitsionas} S., {Whitworth} A.~P., 2002, \mnras, 330, 129

\bibitem[{{Kormendy} \& {Ho}(2013)}]{Kormendy_2013}
{Kormendy} J., {Ho} L.~C., 2013, \araa, 51, 511

\bibitem[{{Kuhlen} \& {Madau}(2005)}]{Kuhlen_2005}
{Kuhlen} M., {Madau} P., 2005, \mnras, 363, 1069

\bibitem[{{Latif} et~al.(2012){Latif}, {Schleicher} \& {Spaans}}]{Latif_2012}
{Latif} M.~A., {Schleicher} D.~R.~G., {Spaans} M., 2012, \aap, 540, A101

\bibitem[{{Latif} et~al.(2013{\natexlab{a}}){Latif}, {Schleicher} \&
  {Schmidt}}]{Latif_2013e}
{Latif} M.~A., {Schleicher} D.~R.~G., {Schmidt} W., 2013{\natexlab{a}}, ArXiv
  e-prints 1310.3680

\bibitem[{{Latif} et~al.(2013{\natexlab{b}}){Latif}, {Schleicher}, {Schmidt} \&
  {Niemeyer}}]{Latif_2013c}
{Latif} M.~A., {Schleicher} D.~R.~G., {Schmidt} W., {Niemeyer} J.,
  2013{\natexlab{b}}, \mnras, 433, 1607

\bibitem[{{Latif} et~al.(2013{\natexlab{c}}){Latif}, {Schleicher}, {Schmidt} \&
  {Niemeyer}}]{Latif_2013a}
{Latif} M.~A., {Schleicher} D.~R.~G., {Schmidt} W., {Niemeyer} J.,
  2013{\natexlab{c}}, \mnras, 430, 588

\bibitem[{{Latif} et~al.(2013{\natexlab{d}}){Latif}, {Schleicher}, {Schmidt} \&
  {Niemeyer}}]{Latif_2013b}
{Latif} M.~A., {Schleicher} D.~R.~G., {Schmidt} W., {Niemeyer} J.,
  2013{\natexlab{d}}, \apjl, 772, L3

\bibitem[{{Latif} et~al.(2013{\natexlab{e}}){Latif}, {Schleicher}, {Schmidt} \&
  {Niemeyer}}]{Latif_2013d}
{Latif} M.~A., {Schleicher} D.~R.~G., {Schmidt} W., {Niemeyer} J.~C.,
  2013{\natexlab{e}}, \mnras, 436, 2989

\bibitem[{{Lodato} \& {Natarajan}(2006)}]{Lodato_2006}
{Lodato} G., {Natarajan} P., 2006, \mnras, 371, 1813

\bibitem[{{Loeb} \& {Rasio}(1994)}]{Loeb_1994}
{Loeb} A., {Rasio} F.~A., 1994, \apj, 432, 52

\bibitem[{{Lynden-Bell}(1969)}]{Lynden-Bell_1969}
{Lynden-Bell} D., 1969, \nat, 223, 690

\bibitem[{{Machacek} et~al.(2003){Machacek}, {Bryan} \& {Abel}}]{Machacek_2003}
{Machacek} M.~E., {Bryan} G.~L., {Abel} T., 2003, \mnras, 338, 273

\bibitem[{{Magorrian} et~al.(1998)}]{Magorrian_1998}
{Magorrian} J. et~al., 1998, \aj, 115, 2285

\bibitem[{{Meece} et~al.(2013){Meece}, {Smith} \& {O'Shea}}]{Meece_2013}
{Meece} G.~R., {Smith} B.~D., {O'Shea} B.~W., 2013, ArXiv e-prints 1309.4090

\bibitem[{{Mesinger} et~al.(2006){Mesinger}, {Bryan} \&
  {Haiman}}]{Mesinger_2006}
{Mesinger} A., {Bryan} G.~L., {Haiman} Z., 2006, \apj, 648, 835

\bibitem[{{Mo} \& {White}(2002)}]{Mo_2002}
{Mo} H.~J., {White} S.~D.~M., 2002, \mnras, 336, 112

\bibitem[{{Mortlock} et~al.(2011)}]{Mortlock_2011}
{Mortlock} D.~J. et~al., 2011, \nat, 474, 616

\bibitem[{{Norman} \& {Bryan}(1999)}]{Norman_1999}
{Norman} M.~L., {Bryan} G.~L., 1999, in S.M. {Miyama}, K.~{Tomisaka},
  T.~{Hanawa}, eds, ASSL Vol. 240: Numerical Astrophysics. pp. 19--+

\bibitem[{{Oh} \& {Haiman}(2002)}]{Oh_2002}
{Oh} S.~P., {Haiman} Z., 2002, \apj, 569, 558

\bibitem[{{Omukai} et~al.(2008){Omukai}, {Schneider} \& {Haiman}}]{Omukai_2008}
{Omukai} K., {Schneider} R., {Haiman} Z., 2008, \apj, 686, 801

\bibitem[{{O'Shea} \& {Norman}(2008)}]{OShea_2008}
{O'Shea} B.~W., {Norman} M.~L., 2008, \apj, 673, 14

\bibitem[{{O'Shea} et~al.(2004){O'Shea}, {Bryan}, {Bordner}, {Norman}, {Abel},
  {Harkness} \& {Kritsuk}}]{OShea_2004}
{O'Shea} B.~W., {Bryan} G., {Bordner} J., {Norman} M.~L., {Abel} T., {Harkness}
  R., {Kritsuk} A., 2004, ArXiv Astrophysics e-prints 0403044

\bibitem[{{Palla} et~al.(1983){Palla}, {Salpeter} \& {Stahler}}]{Palla_1983}
{Palla} F., {Salpeter} E.~E., {Stahler} S.~W., 1983, \apj, 271, 632

\bibitem[{{Pawlik} \& {Schaye}(2009)}]{Pawlik_2009}
{Pawlik} A.~H., {Schaye} J., 2009, \mnras, 396, L46

\bibitem[{{Planck Collaboration} et~al.(2013)}]{Planck_2013a}
{Planck Collaboration} et~al., 2013, ArXiv e-prints 1303.5076

\bibitem[{{Press} \& {Schechter}(1974)}]{ps74}
{Press} W.~H., {Schechter} P., 1974, 187, 425

\bibitem[{{Prieto} et~al.(2013){Prieto}, {Jimenez} \& {Haiman}}]{Prieto_2013}
{Prieto} J., {Jimenez} R., {Haiman} Z., 2013, \mnras, 436, 2301

\bibitem[{{Regan} \& {Haehnelt}(2009{\natexlab{a}})}]{Regan_2009b}
{Regan} J.~A., {Haehnelt} M.~G., 2009{\natexlab{a}}, \mnras, 396, 343

\bibitem[{{Regan} \& {Haehnelt}(2009{\natexlab{b}})}]{Regan_2009}
{Regan} J.~A., {Haehnelt} M.~G., 2009{\natexlab{b}}, \mnras, 393, 858

\bibitem[{{Salpeter}(1964)}]{Salpeter_1964}
{Salpeter} E.~E., 1964, \apj, 140, 796

\bibitem[{{Schleicher} et~al.(2010){Schleicher}, {Spaans} \&
  {Glover}}]{Schleicher_2010}
{Schleicher} D.~R.~G., {Spaans} M., {Glover} S.~C.~O., 2010, \apjl, 712, L69

\bibitem[{{Shang} et~al.(2010){Shang}, {Bryan} \& {Haiman}}]{Shang_2010}
{Shang} C., {Bryan} G.~L., {Haiman} Z., 2010, \mnras, 402, 1249

\bibitem[{{Shlosman} et~al.(1989){Shlosman}, {Frank} \&
  {Begelman}}]{Shlosman_1989}
{Shlosman} I., {Frank} J., {Begelman} M.~C., 1989, \nat, 338, 45

\bibitem[{{Sijacki} \& {Springel}(2006)}]{Sijacki_2006}
{Sijacki} D., {Springel} V., 2006, \mnras, 366, 397

\bibitem[{{Sijacki} et~al.(2007){Sijacki}, {Springel}, {Di Matteo} \&
  {Hernquist}}]{Sijacki_2007}
{Sijacki} D., {Springel} V., {Di Matteo} T., {Hernquist} L., 2007, \mnras, 380,
  877

\bibitem[{{Sijacki} et~al.(2009){Sijacki}, {Springel} \&
  {Haehnelt}}]{Sijacki_2009}
{Sijacki} D., {Springel} V., {Haehnelt} M.~G., 2009, \mnras, 400, 100

\bibitem[{{Smith} et~al.(2009){Smith}, {Turk}, {Sigurdsson}, {O'Shea} \&
  {Norman}}]{Smith_2009}
{Smith} B.~D., {Turk} M.~J., {Sigurdsson} S., {O'Shea} B.~W., {Norman} M.~L.,
  2009, \apj, 691, 441

\bibitem[{{Spaans} \& {Silk}(2006)}]{Spaans_2006}
{Spaans} M., {Silk} J., 2006, \apj, 652, 902

\bibitem[{{Stacy} et~al.(2012){Stacy}, {Greif} \& {Bromm}}]{Stacy_2012}
{Stacy} A., {Greif} T.~H., {Bromm} V., 2012, \mnras, 422, 290

\bibitem[{{Tanaka} \& {Li}(2013)}]{Tanaka_2013}
{Tanaka} T.~L., {Li} M., 2013, ArXiv e-prints 1310.0859

\bibitem[{{The Enzo Collaboration} et~al.(2013)}]{Enzo_2013}
{The Enzo Collaboration}, {Bryan} G.~L., {Norman} M.~L., {O'Shea} B.~W., {Abel}
  T., {Wise} J.~H., {Turk} M.~J., et~al., 2013, ArXiv e-prints 1307.2265

\bibitem[{{Toomre}(1964)}]{Toomre_1964}
{Toomre} A., 1964, \apj, 139, 1217

\bibitem[{{Truelove} et~al.(1997){Truelove}, {Klein}, {McKee}, {Holliman},
  {Howell} \& {Greenough}}]{Truelove_1997}
{Truelove} J.~K., {Klein} R.~I., {McKee} C.~F., {Holliman} II J.~H., {Howell}
  L.~H., {Greenough} J.~A., 1997, \apjl, 489, L179+

\bibitem[{{Turk} et~al.(2009){Turk}, {Abel} \& {O'Shea}}]{Turk_2009}
{Turk} M.~J., {Abel} T., {O'Shea} B., 2009, Science, 325, 601

\bibitem[{{Turk} et~al.(2011){Turk}, {Smith}, {Oishi}, {Skory}, {Skillman},
  {Abel} \& {Norman}}]{YT}
{Turk} M.~J., {Smith} B.~D., {Oishi} J.~S., {Skory} S., {Skillman} S.~W.,
  {Abel} T., {Norman} M.~L., 2011, \apjs, 192, 9

\bibitem[{{Turk} et~al.(2012){Turk}, {Oishi}, {Abel} \& {Bryan}}]{Turk_2012}
{Turk} M.~J., {Oishi} J.~S., {Abel} T., {Bryan} G.~L., 2012, \apj, 745, 154

\bibitem[{{Van Borm} \& {Spaans}(2013)}]{VanBorm_2013}
{Van Borm} C., {Spaans} M., 2013, \aap, 553, L9

\bibitem[{{Venemans} et~al.(2013)}]{Venemans_2013}
{Venemans} B.~P. et~al., 2013, \apj, 779, 24

\bibitem[{{Volonteri}(2010)}]{Volonteri_2010a}
{Volonteri} M., 2010, \aapr, 18, 279

\bibitem[{{Volonteri} \& {Begelman}(2010)}]{Volonteri_2010}
{Volonteri} M., {Begelman} M.~C., 2010, \mnras, 409, 1022

\bibitem[{{Wise} et~al.(2008){Wise}, {Turk} \& {Abel}}]{Wise_2008}
{Wise} J.~H., {Turk} M.~J., {Abel} T., 2008, \apj, 682, 745

\bibitem[{{Wolcott-Green} et~al.(2011){Wolcott-Green}, {Haiman} \&
  {Bryan}}]{Wolcott-Green_2011}
{Wolcott-Green} J., {Haiman} Z., {Bryan} G.~L., 2011, \mnras, 418, 838

\bibitem[{{Zel'Dovich} \& {Novikov}(1964)}]{Zeldovich_1964}
{Zel'Dovich} Y.~B., {Novikov} I.~D., 1964, Soviet Physics Doklady, 9, 246

\end{thebibliography}

\appendixtitleon
\appendixtitletocon
\begin{appendices}
\section{The Probability Distribution of the Initial  Angular Momentum
  and its Dependence on Halo Rareness and Environment} 
\label{appendixA}
\begin{figure*}
  \centering 
  \begin{minipage}{175mm}      \begin{center}
    \centerline{
      \psfig{file=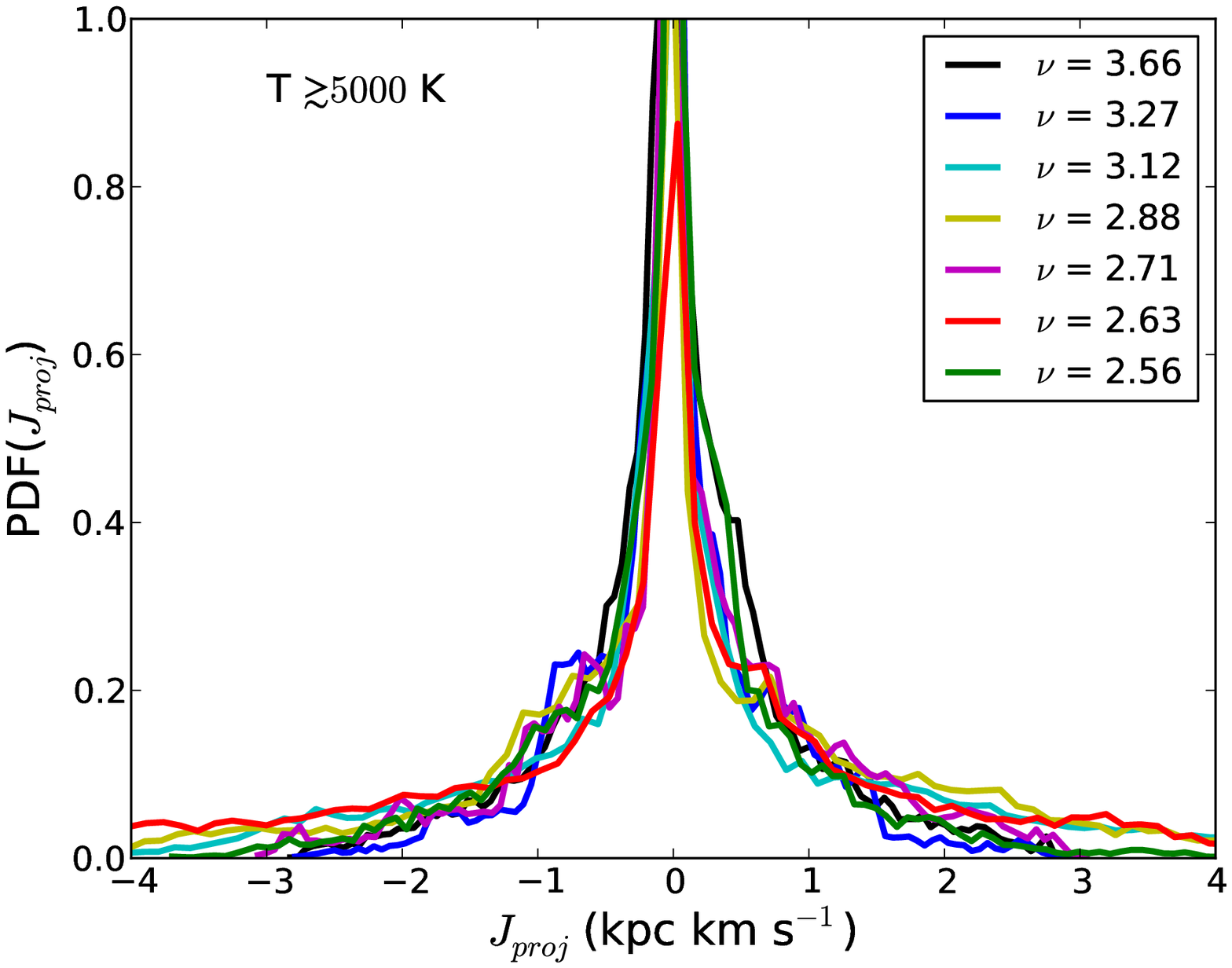,width=9cm}
      \psfig{file=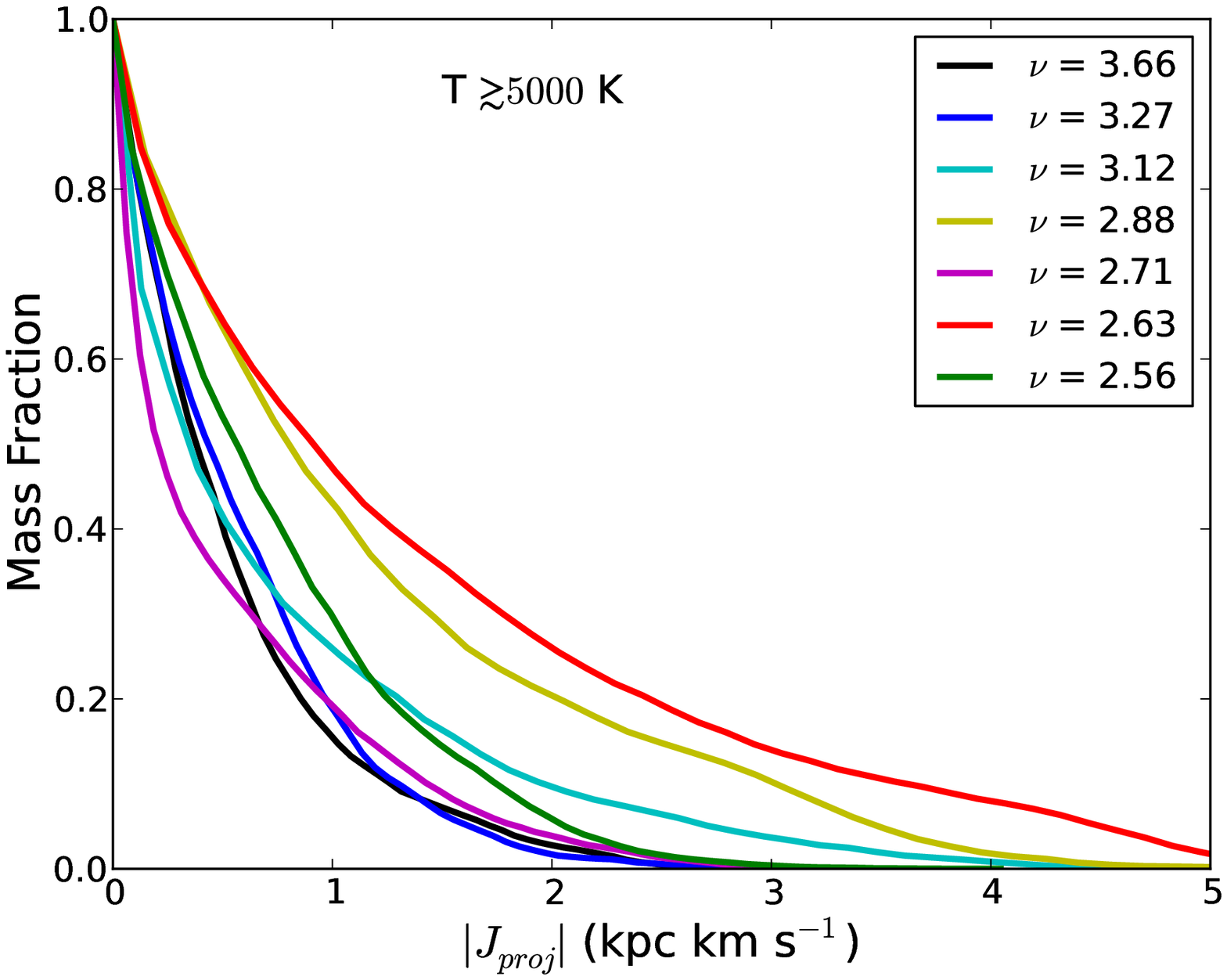,width=9cm}}
    \caption[]{\label{5000K}
      {\it Left Panel:} 
      The Probability Distribution Function of the projected angular momentum 
      for the gas in the halo when the virial temperature is $\sim$
      5000 K. The PDF is strongly peaked at zero as expected with tails to higher 
      values. The line colours match those in previous plots, but are reordered 
      to be in descending order of $\nu$. 
      {\it Right Panel:}
      The Mass Fraction of gas with an absolute value of projected angular 
      momentum above a given value. }
   \end{center} \end{minipage}
\end{figure*}
 
\begin{figure*}
  \centering 
  \begin{minipage}{175mm}      \begin{center}
    \centerline{
      \psfig{file=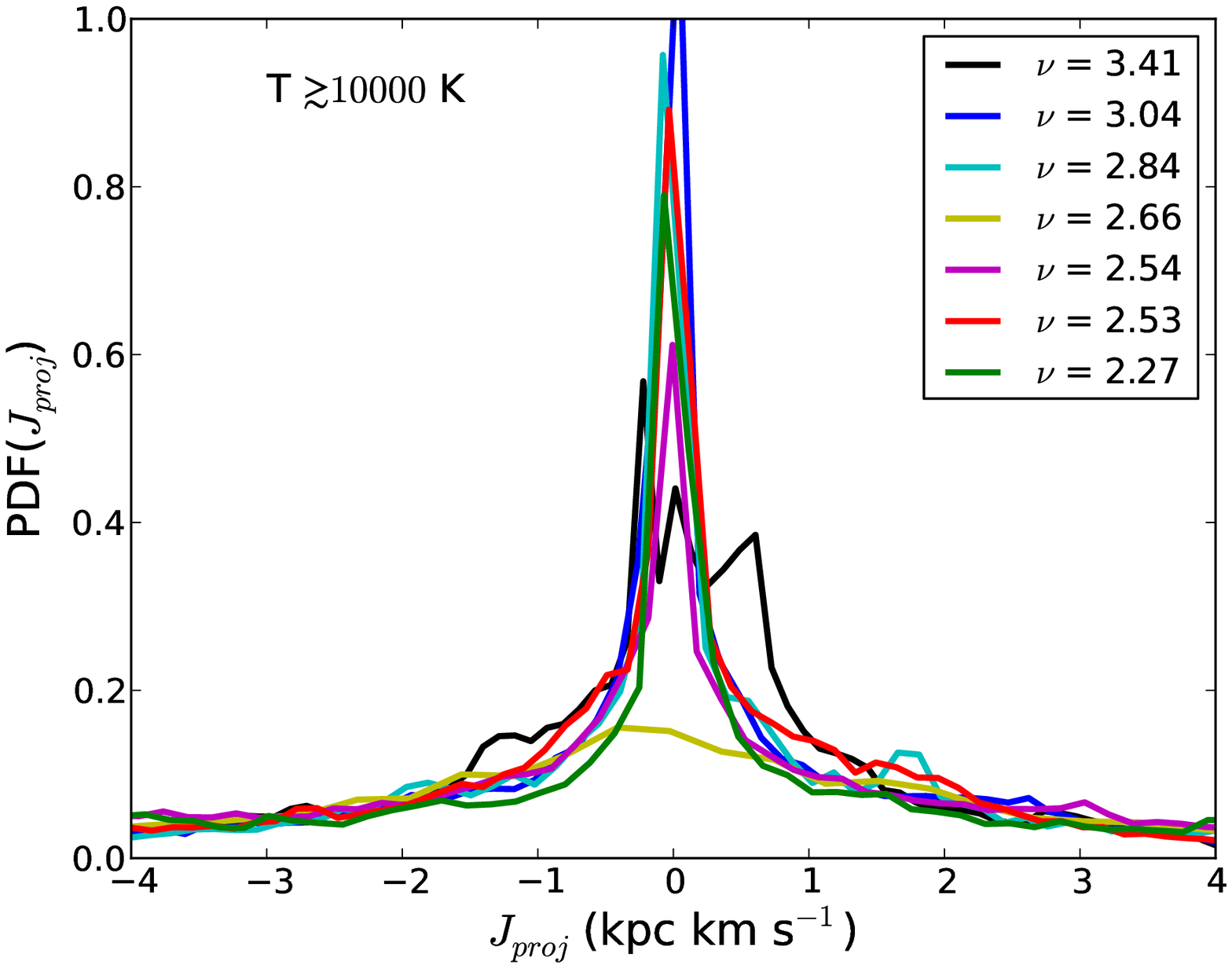,width=9cm}
      \psfig{file=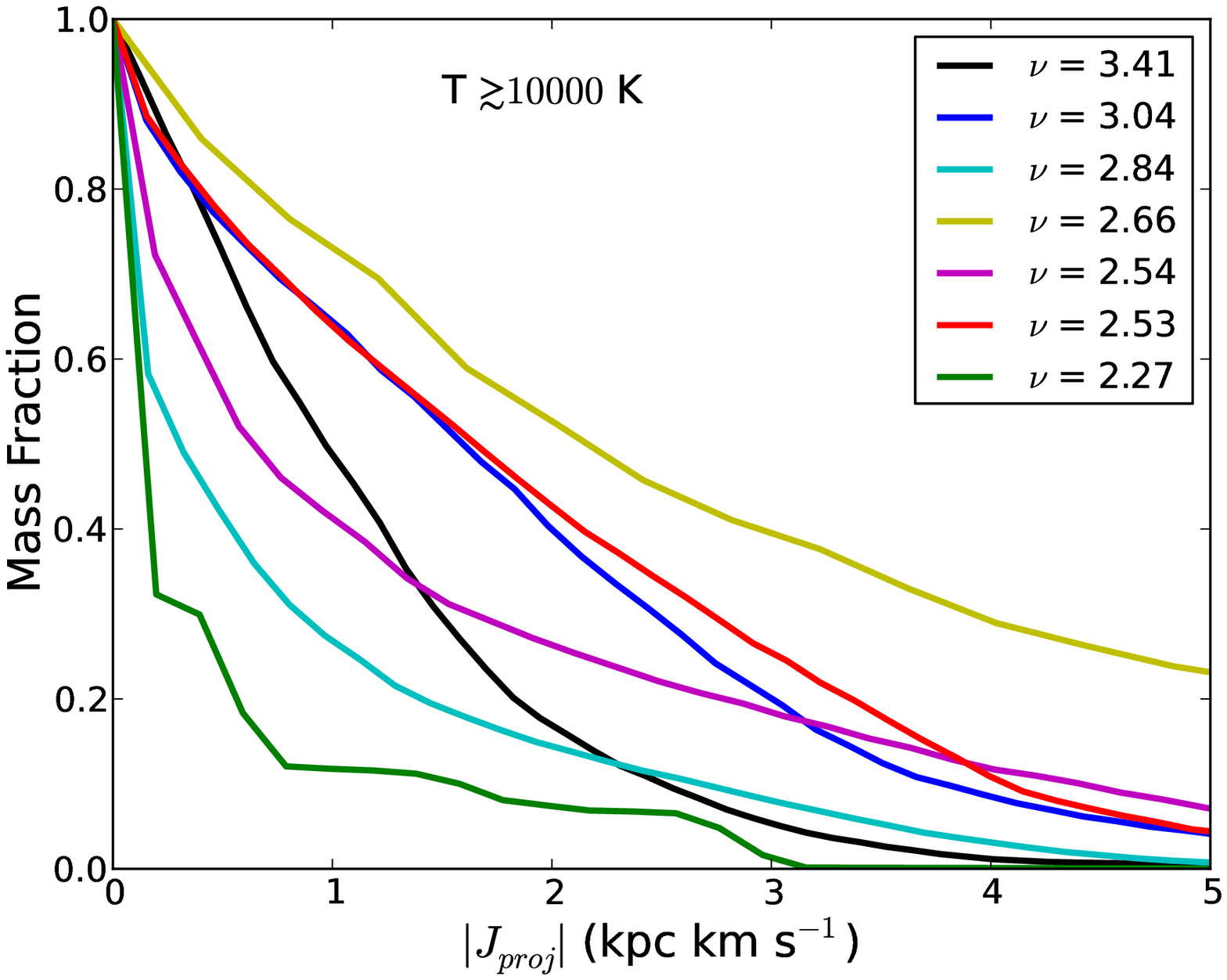,width=9cm}}
    \caption[]{\label{10000K}
      {\it Left Panel:} 
      The Probability Distribution Function of the projected angular momentum 
      for the gas in the halo at the end of the simulation when the virial 
      temperature is $\gtrsim$ 10000 K. The PDF is strongly peaked at zero 
      as expected with tails to higher values. The line colours match those in 
      previous plots, but are reordered in descending order of $\nu$. 
      {\it Right Panel:}
      The Mass Fraction of gas with an absolute value of projected angular 
      momentum above a given value. }
   \end{center} \end{minipage}
\end{figure*}
 
\begin{figure*}
  \centering    
    \begin{center}
      \centerline{
        \includegraphics[width=9cm]{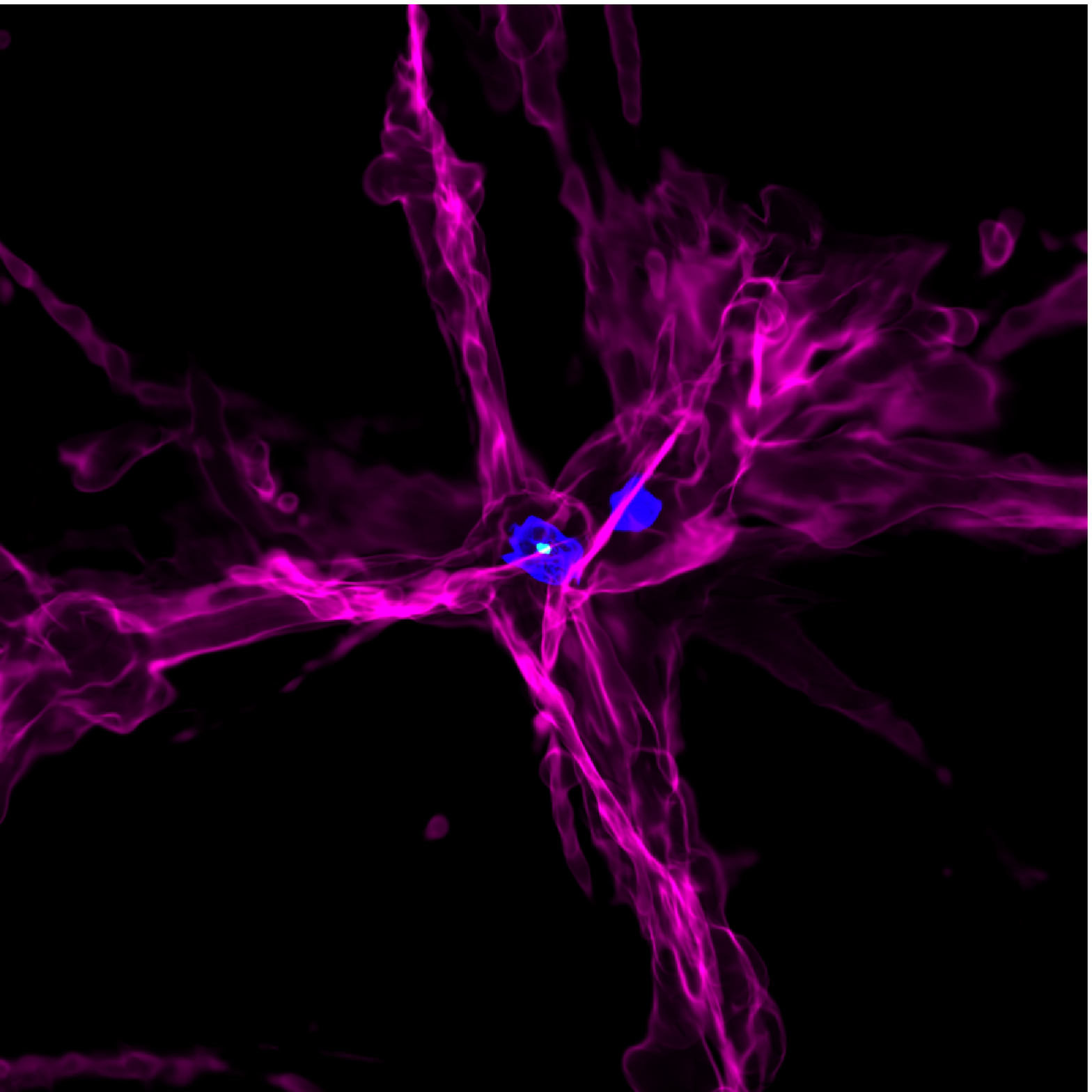}
        \includegraphics[width=9cm]{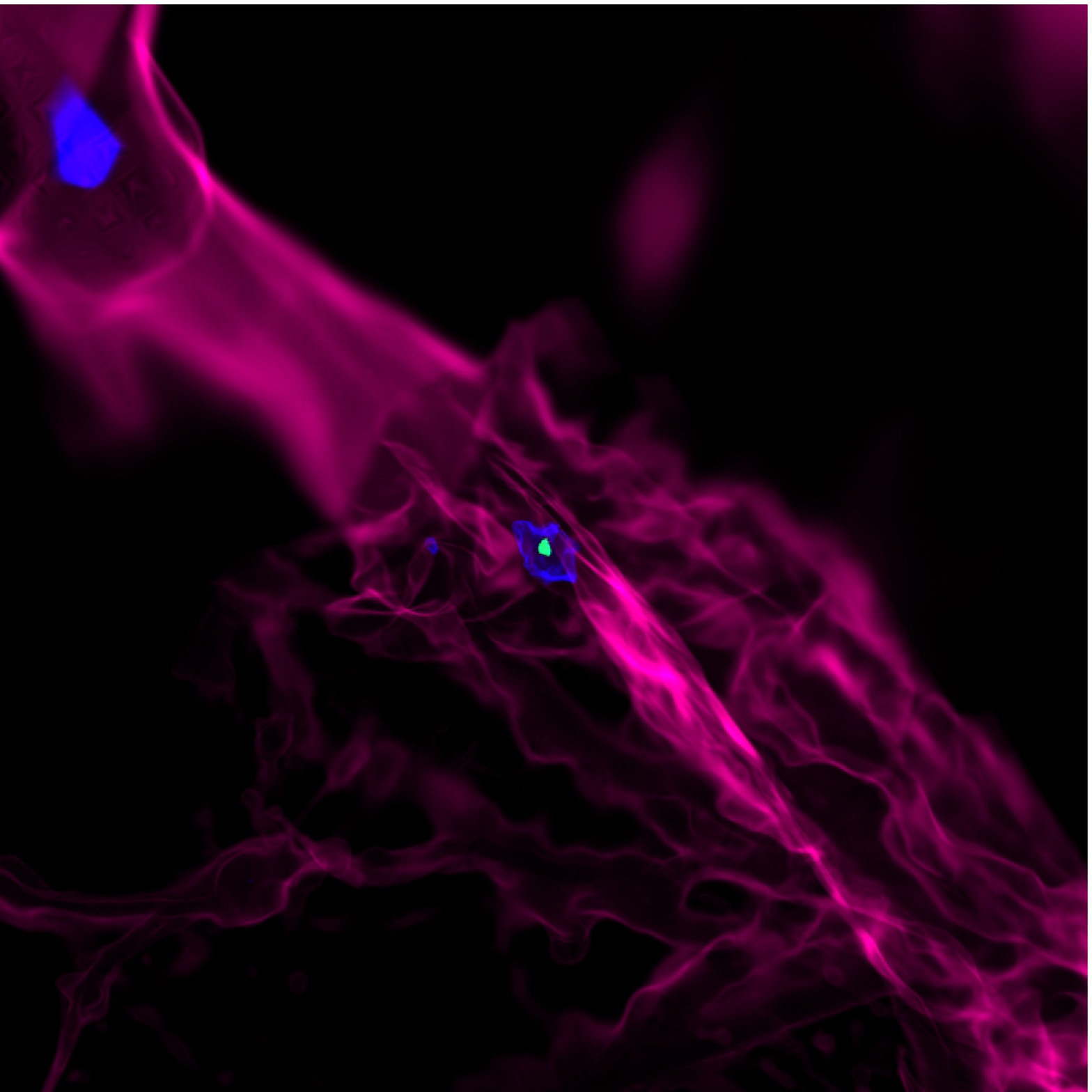}}
      \caption[]
      {\label{Filaments}
        {\it Left Panel:} 
        The cosmic web centred on the densest point in simulation A. The
        volume rendering is $\sim 1.5$ kpc on the side. The collapsing object
        at the centre is shown in cyan. 
        {\it Right Panel:}
        Volume rendered visualisation of simulation B. The densest point and 
        centre of the disk is found midway along a filament. Other over-densities
        are also clearly visible, shown in blue. The width of the plot is 
        $\sim 3 $ kpc on the side. 
      }   
    \end{center}
\end{figure*}


As discussed in the introduction the space density of massive seed black holes required to 
explain the presence of the observed billion solar mass black holes at very high redshift is 
very much smaller than the space density of DM haloes at the atomic cooling threshold.  
Low metallicity and a high amplitude of the Lyman-Werner radiation may therefore not necessarily 
be the only criteria required for the (efficient) formation of a massive seed black hole. 
As we were able to perform here a larger sample of simulations of these haloes compared to our 
previous work \citep{Regan_2009} we have looked also briefly into the possibility of 
systematic trends with the ``rareness'' and the environment of the simulated haloes 
\citep[see also][]{Prieto_2013}.  We thereby characterise the ``rareness'' of the halo 
(or peak height) in terms of the RMS fluctuation amplitude of the linearly extrapolated 
density field  
\begin{equation}
\nu = {\delta_c \over \sigma(M) D(z)},
\end{equation}  
where $\delta_c$ is the threshold over-density from \cite{ps74}, $D(z)$ is the growth factor
and $\sigma(M)$ is the mass fluctuation inside a halo of mass M: 
\begin{equation}
\sigma(M)^2 = \int {k^2 \over {2 \pi^2}} P(k) W(kR) dk
\end{equation}  
\noindent where the integral is over the wavenumber $k$, $P(k)$ is the power spectrum and 
$W(kR)$ is the top hat window function. As the haloes undergo collapse, it is the lowest 
angular momentum gas that can fall radially to the centre \citep{Dubois_2012, Bellovary_2013}.
As the collapse goes on angular momentum redistribution and cancellation occurs 
\citep{Wise_2008, Regan_2009}. In order to increase the range of $\nu$ beyond that of 
our fiducial simulations we ran an extra 150 extra dark matter simulations to find a rarer 
peak than in our six fiducial simulations. We  picked the highest $\nu$ peak from these 150 runs 
and reran the simulation at the same maximum refinement level as the fiducial simulations. 
This high-$\nu$ halo is included in the following results. \\ \indent
Figure \ref{5000K} shows the PDF of the projected angular momentum for our seven simulated haloes 
with virial temperatures of $\sim 5000$ K. The projected angular momentum is calculated by 
first rotating into the coordinate system defined by the inertia tensor. The left panel of 
Figure \ref{5000K} shows the PDF for each halo when it has reached a virial temperature of 
$\rm{T_{vir}} \sim 5000$ K. At  $\rm{T_{vir}} \sim 5000$ K  the halo is still growing by mass 
inflow along the surrounding filamentary structures and is just about to collapse. The 
angular momentum distribution at this stage should be a good representation of the 
initial angular momentum distribution with which the gas enters into the collapse.
Figure \ref{5000K} shows no clear trend of lower angular momentum gas with rareness of the  halo. 
The angular momentum distributions are pretty symmetrically distributed at zero with similar 
amounts of co- and counter-rotating gas which explains why the gas ``looses'' 
so efficiently angular momentum during the collapse due to angular momentum cancellation. 
There is also no evidence of a trend towards more symmetric distribution in rarer haloes as 
suggested by \cite{Dubois_2012}, but the range of $\nu$ of our haloes is still rather moderate 
due to the 
small size of the  simulation box.  Unfortunately, with our current set-up of the simulations 
it is not straightforward to extend the range further.  The right hand panel of Figure 
\ref{5000K} shows the mass fraction of absolute projected angular momentum above a given value. 
Similarly no trend with the rareness of the halos is seen.  Figure \ref{10000K} shows the PDF 
and mass fraction plots for the outputs at the end of our simulation runs 
(i.e. when $\rm{T_{vir}} \gtrsim 10000$ K). Again no clear trend with $\nu$ is detected. Other 
properties of the seven haloes we have looked at show likewise no significant trend. \\ \indent
In Figure \ref{Filaments} we illustrate the location of the central 
object amid the cosmic web for two of the simulations. The left panel of Figure \ref{Filaments}
shows the central object in simulation A while the right hand panel shows the 
location of the central object in simulation B. The object in simulation A is located at the 
knot of a cosmic web of gas while the central object in simulation B is located midway 
along a filament. Both objects therefore have formed in relatively different environments with 
regard to gas inflow and in particular filamentary inflow. The visualisations also show the 
presence of other dense clumps of gas forming within $\lesssim 5$ kpc of the central object in 
each simulation. We are not able to confirm the trends in different collapse properties of 
haloes at the atomic cooling threshold found by Prieto et al. for our (admittedly smaller) sample 
of simulations. The difference in resolution between the studies is, however, significant. 
The study undertaken by \cite{Prieto_2013} used much larger box sizes (up to 8 Mpc$^3$), but with
reduced resolution (8 pc at maximum). Therefore, they were able to study the dynamical 
evolution of the central object over much longer periods, but were not able to resolve the 
sub parsec evolution of these objects.

\end{appendices}
\end{document}